\documentclass[aps,pra,twocolumn,amsmath,nofootinbib]{revtex4}
\usepackage[dvips]{color}
\usepackage[normalem]{ulem}

\definecolor{g-blue}{rgb}{0.83,0.95,1}
\definecolor{g-yellow}{rgb}{1,1,0.7}
\definecolor{g-green}{rgb}{0.9,1,0.9}
\definecolor{green}{rgb}{0,0.6,0}
\definecolor{cyan}{rgb}{0,0.7,0.7}
\definecolor{black}{rgb}{0,0,0}
\definecolor{grey}{rgb}{0.4 ,0.4 ,0.4 }

\usepackage{bm}
\usepackage{amssymb}
\usepackage{amsfonts}
\usepackage{amsmath}
\usepackage{graphicx}
\usepackage{amsmath,bm,epsfig}
\usepackage[alsoload=astro]{siunitx}
\usepackage{xcolor}
\def \ed {\end{document}}

\newcommand{\Eq}[1]{Eq.\,(\ref{#1})}
\newcommand{\Eqs}[1]{Eqs.\,(\ref{#1})}
\newcommand{\Fig}[1]{Fig.\,\ref{#1}}
\newcommand{\Sec}[1]{Sec.\,\ref{#1}}
\newcommand{\Secs}[1]{Secs.\,\ref{#1}}

\def\be{\begin{equation}}
\def\ee{\end{equation}}
\def\bea{\begin{eqnarray}}
\def\eea{\end{eqnarray}}
\def\bse{\begin{subequations}}
\def\ese{\end{subequations}}
\newcommand{\BE}[1] {\begin{eqnarray}\label{#1}}

\let \= \equiv  \let\*\cdot 

 \def\1{\bm1} 

\def\<{\left\langle}    \def\>{\right\rangle}
\def\({\left(}          \def\){\right)}
\def \[ {\left [} \def \] {\right ]}

\def\He4 {$^4$He~}

\begin{document}

\title{ Wave Turbulence in Self-Gravitating Bose Gases and Nonlocal Nonlinear Optics }

\author{Jonathan Skipp}
\affiliation{Centre for Complexity Science, University of Wawick, Coventry CV4 7AL, UK}

\author{Victor L'vov}
\affiliation{Department of Chemical and Biological Physics, Weizmann Institute of Science, Rehovot 76100, Israel }
  
\author{Sergey Nazarenko}
\affiliation{Universit\'{e}  C\^{o}te d'Azur,   CNRS, Institut de Physique de Nice, Parc Valrose, 06108 Nice, France
}

\begin{abstract} 
 We develop the theory of weak wave turbulence in systems described by the Schr{\"o}dinger-Helmholtz equations in two and three dimensions. This model contains as limits both the familiar cubic nonlinear Schr{\"o}dinger equation, and the Schr{\"o}dinger-Newton equations. 
    		The latter, in three dimensions, are a nonrelativistic model of fuzzy dark matter which has a nonlocal gravitational self-potential, and in two dimensions they describe nonlocal nonlinear optics in the paraxial approximation.
 We show that in the weakly nonlinear limit the Schr{\"o}dinger-Helmholtz equations have a simultaneous inverse cascade of particles and a forward cascade of energy. We interpret the inverse cascade as a nonequilibrium condensation process, which is a precursor to structure formation at large scales (for example the formation of galactic dark matter haloes or optical solitons).
We show that for the Schr{\"o}dinger-Newton equations in two and three dimensions, and in the two-dimensional nonlinear Schr{\"o}dinger equation, the particle and energy fluxes are carried by small deviations from thermodynamic distributions, rather than the Kolmogorov-Zakharov cascades that are familiar in wave turbulence. We develop a differential approximation model to characterise such ``warm cascade'' states.
\end{abstract}

\maketitle


\section{\label{s:intro} Introduction}

\subsection{\label{ss:WT} Wave turbulence cascades}

The dynamical and statistical behaviour of random weakly-interacting waves is responsible for many important physical effects across  applications ranging from quantum to classical and to astrophysical scales \citep{zakharov1992kolmogorovbook,nazarenko2011waveturbbook}. Assuming weak nonlinearity and random phases, such behaviour is described by the theory of weak wave turbulence~\citep{zakharov1992kolmogorovbook,nazarenko2011waveturbbook}. As in the theory of classical hydrodynamic turbulence, weak wave turbulence theory can predict nonequilibrium statistical states characterised by cascades of energy and/or other invariants through scales. Sometimes, similarly to 2D classical turbulence, such cascades are dual, with one invariant cascading to small scales (direct cascade) and the other invariant towards large scales (inverse cascade). An inverse cascade often leads to accumulation of the turbulence spectrum near the largest scale of the system, which is analogous to Bose-Einstein condensation. Large-scale coherent structures may form out of such a condensate and further evolve via mutual interactions and interactions with the background of random waves, thereby realising a scenario of order emerging from chaos. 

In the present paper, we will study a precursor to such a process of coherent structure formation by developing the wave turbulence theory and describing the dual cascade in the so-called ``Schr{\"o}dinger-Helmholtz equations'' {that arise in cosmological and nonlinear optics applications.

\subsection{\label{ss:SHE}Schr{\"o}dinger-Helmholtz equations}
The Schr{\"o}dinger-Helmholtz equations are the nonlinear partial differential equations
\begin{subequations}\label{SHE}
	\begin{align}
 		i \partial_t \psi  +  \nabla^2   \psi  -  V[\psi] \psi &= 0\,,\label{SHE:NLS}\\
 		\nabla^2 V - \Lambda V &= \gamma \lvert\psi \rvert^2\label{SHE:Helmholtz}	   
 	\end{align}
\end{subequations}
for a complex scalar field $\psi(\mathbf{x},t)$ in which $V[\psi]$ plays the role of (potential) interaction energy and $\Lambda$ and $\gamma$ are constants. We will be interested in systems set in three and two spatial dimensions (3D and 2D, respectively).
 
 Before proceeding in the body of the paper with developing the statistical description of the nonlinear field $\psi$ in the framework of \Eqs{SHE}, we will first outline in this \Sec{ss:SHE} the important physical contexts in which \Eqs{SHE} have been used, the previous results found, and the findings that we anticipate will arise from our approach. 
 
 Notice that depending on the spatial scale of interest $\ell$,  one term or the other on the left-hand side of \Eq{SHE:Helmholtz} is dominant. For $ \ell \gg \ell_*= 1/\sqrt{\Lambda}$ the Schr{\"o}dinger-Helmholtz equations~\eqref{SHE} become the more familiar cubic nonlinear Schr{\"o}dinger equation, discussed in \Sec{sss:NLSE},  while for $ \ell \ll \ell_* $ they turn into the Schr{\"o}dinger-Newton equations, see \Sec{sss:SNE}.  Finally, in  \Sec{sss:SHE} we return to interpret the Schr{\"o}dinger-Helmholtz \Eqs{SHE}  in light of the discussion of these limits. 
    
\subsubsection{\label{sss:NLSE}Large-scale limit: The nonlinear Schr{\"o}dinger equation}

In the limit of large scales, $\ell \gg \ell_*$,  the first term on the left-hand side of \Eq{SHE:Helmholtz} can be neglected and one immediately finds that $V[\psi]=- (\gamma/\Lambda) |\psi|^2$. The constant  $\gamma/\Lambda$  can be removed by proper renormalization of $|\psi|^2$, leaving only the sign of this constant, denoted as $s=\pm 1$. Thus the Schr{\"o}dinger-Helmholtz \Eqs{SHE} become the nonlinear Schr{\"o}dinger equation 
 \begin{equation} \label{NLSE}
 i \partial_t \psi  +  \nabla^2   \psi   + s \lvert \psi \rvert^2 \psi =0\,,
 \end{equation}
  also known as the Gross-Pitaevskii equation~\citep{pitaevskiistringari2016book}. This equation has a cubic, spatially local, attractive (for $s=+1$) or  repulsive (for $s=-1$)  interaction.

The nonlinear Schr{\"o}dinger\, \Eq{NLSE} is well known in the study of Bose-Einstein condensates~\citep{pitaevskiistringari2016book},
where $\psi$ is the wavefunction of a system of identical bosons in the Hartree-Fock approximation~\citep{gross_structure_1961, pitaevskii_vortex_1961} and the nonlinearity is due to $s$-wave scattering.  
		(As well as normalising the coupling constant $s$ to $\pm 1$, units are further chosen such that the reduced Planck constant $\hbar=1$ and the boson mass $m=1/2$.)

Equation \eqref{NLSE} is also familiar in the field of nonlinear optics~\citep{newellmoloney1992nonlinearopticsbook, boyd2008nonlinearopticsbook} when a light beam, whose electric field is slowly modulated by an envelope $\psi$ (such that its intensity is $\lvert\psi\rvert^2$), impinges on a dispersive, nonlinear medium, inducing a nonlinear change in the medium's refractive index via the Kerr effect. 
Equation~\eqref{NLSE} then describes the evolution of the beam's envelope in the paraxial approximation, where $t$ becomes the length along the beam axis, and the remaining spatial directions are transverse to the beam.
		(In the optics application units are chosen such that $k_0 n_0=1/2$  
		where $k_0$ is the free space wavenumber of the input beam and $n_0$ is the refractive index of the medium, 
		normalising the coefficient of the Laplacian term to unity.)

In this context $s$ is the normalised Kerr coefficient, and the cases with $s=+1$ or $-1$ are known as the focusing or defocusing nonlinear Schr{\"o}dinger equation respectively, terminology that we adopt here in the general case.

The nonlinear Schr{\"o}dinger \Eq{NLSE}   is studied in a great many other systems due to its universality in describing the slowly-varying envelope of a monochromatic wave in a weakly nonlinear medium~\citep{whitham1999waves}. We shall not pursue its other applications in this work, instead merely noting that due to its universality many monographs and papers have been dedicated to the study of \Eq{NLSE} and its solutions.

\subsubsection{\label{sss:SNE} Small-scale limit: The Schr{\"o}dinger-Newton equations}

Now we focus on scales  $\ell  \ll \ell_*$, when the second term in the left-hand side of \Eq{SHE:Helmholtz} dominates. Then the Schr{\"o}dinger-Helmholtz \Eqs{SHE} simplify to the coupled equations
\begin{subequations}\label{SNE}
    \begin{align}
            i \partial_t \psi  +  \nabla^2   \psi  -  V[\psi] \psi &= 0\,,\label{SNE:NLS}\\
       \nabla^2 V &= \gamma \lvert\psi \rvert^2\,.\label{SNE:Poisson}	   
    \end{align}
  \end{subequations}
	In three dimensions
if we retain the interpretation of $\psi(\mathbf{x},t)$ as a boson wavefunction, we see that the nonlinearity in \Eq{SNE:NLS} is nonlocal, coming from an extended potential $V[\psi]$ that solves the Poisson \Eq{SNE:Poisson} for which the source is proportional to the boson number density $\rho = \lvert \psi\rvert^2$. Specifying 
		$\gamma=\pi$, and noting that we have chosen units in which $\hbar=1$, $m=1/2$, and Newton's gravitational constant $G=1$,
we observe that \Eqs{SNE} describe a dilute Bose gas moving at nonrelativistic speeds under the influence of a Newtonian gravitational potential generated by the bosons themselves. It is for this reason that \Eqs{SNE} are known as the Schr{\"o}dinger-Newton equations. 
	(The derivation of \Eqs{SNE} from a Klein-Gordon action with a general relativistic metric can be found in the literature, 
	for example~\citep{marsh2016axion, niemeyer2020small}.)

The use of \Eqs{SNE} to represent self-gravitating Bose gases in the Newtonian limit is important in cosmology, where they are used to model ``fuzzy dark matter''. This is the hypothesis that dark matter is comprised of ultra-light ($m\lesssim \SI{1e-22}{\electronvolt}$) scalar bosons whose de Broglie wavelengths are on the order of galaxies ($\lambda_\mathrm{dB}\sim \SI{1}{\kilo\parsec}$)~\citep{HuFuzzy2000, SuarezBECDM2014, Chavanis2015_SelfGravBECs, marsh2016axion, hui2017ultralight, ferreira2020ultra}.
 In this scenario galactic dark matter haloes are gigantic condensates of this fundamental boson, trapped by their own gravity and supported by quantum pressure arising from the uncertainty principle~\citep{HuFuzzy2000, baldeschi1983massive, lee1996galactic, Chavanis2015_SelfGravBECs, schive2014cosmic, verma2019formation, niemeyer2020small}.

Fuzzy dark matter is an alternative to the standard model of cosmology which supposes that dark matter is comprised of thermal but sub-luminal, weakly interacting massive particles, i.e.,\ ``cold dark matter''~\citep{OverduinWesson_WIMPs_2004}.
	While cold dark matter is successful at describing the observed large-scale structure of the universe, its accelerated expansion, and the fluctuations of the cosmic microwave background~\citep{PeeblesRatra_CDM_2003, Springel_CDM_2005}, at small scales it fails to reconcile observations with cosmological simulations, particularly in matching the inferred flat density profiles of galactic dark matter haloes with the cuspy profiles found in simulations, and the lack of observed satellite dwarf galaxies as compared to theoretical predictions~\citep{Weinberg_SmallScaleControv_2015,Bullock_SmallScaleControv_2017}. 
	By contrast, in fuzzy dark matter  
		galactic cores arise naturally as compact soliton-like objects structures with core radii on the order of $\lambda_\mathrm{dB}$, below which fine structure is suppressed by the uncertainty principle~\citep{harko2011bose, marsh2015axion}
and, when included in the model, $s$-wave scattering~\citep{Chavanis2015_SelfGravBECs,lee1996galactic,colpi1986boson}, providing a resolution to the small-scale problems of cold dark matter.  At large scales the two models become indistinguishable~\citep{schive2014cosmic}. Thus, until the precise nature of dark matter particles is identified, fuzzy dark matter must be considered alongside cold dark matter when investigating the formation of large-scale structure in the early universe~\citep{schive2014cosmic, mocz2017galaxyI, mocz2019first, mocz2019galaxyII}.

	Like the nonlinear Schr{\"o}dinger\, \Eq{NLSE}, the Schr{\"o}dinger-Newton \Eqs{SNE} also have applications in nonlinear optics. Here~\eqref{SNE:NLS} is again the equation for the envelope of the beam
		in two transverse spatial dimensions and the distance along the beam is again the time-like dimension.	 
		$V[\psi]$ is now the change in refractive index 	of the optical sample induced by the incident beam, whose nonlocality is expressed in 
		\Eq{SNE:Poisson}. 
		This can be due to the refractive index being temperature-dependent 
		and \eqref{SNE:Poisson} describing the diffusion of the incident beam energy through the medium as heat:
		the thermo-optic effect~\citep{boyd2008nonlinearopticsbook, castillo1996formation}. 
		Alternatively, in nematic liquid crystals the refractive index depends on the orientation of the liquid crystal molecules 
		with respect to the wavevector of the incident beam, and \eqref{SNE:Poisson} describes the reorientation 
		induced by the electric (or magnetic) field of the beam, 
		which diffuses through the sample due to long-range elastic interactions between the molecules~\citep{khoo2007liquidcrystalsbook}.
		
	Nonlocal nonlinear optics manifest many phenomena that are the nonlocal versions of the equivalent local phenomenon, 
	for example (but by no means limited to) 
	solitons~\citep{snyder1997accessible, castillo1996formation, conti2003route, rotschild2005ellipticsolitons}, 
	soliton interactions~\citep{rotschild2006long}, 
	modulational instability and collapse~\citep{perez2000dynamics, bang2002collapse, peccianti2003optical}, and
	shocks and shock turbulence~\citep{ghofraniha2007shocks, xu2015coherent}.
	In addition, comparisons can be made between nonlinear optical systems and fuzzy dark matter 
	by virtue of \Eqs{SNE} describing them both. 
	Indeed recent optics experiments~\citep{Faccio2016_OpticalNewtSchro, Beckenstein2015_OpticalNewtSchro},
	and theoretical works~\citep{navarrete2017spatial, paredes2020optics} have drawn direct analogies 
	between optical systems that can be realised in the laboratory and astrophysical systems on the scale of galaxies.

\subsubsection{\label{sss:SHE} Physical applications of the Schr{\"o}dinger-Helmholtz equations}

The Schr{\"o}dinger-Helmholtz \Eqs{SHE}, then, are a model that captures the physics present in both \Eq{NLSE} and \Eqs{SNE}. 
		Applied to fuzzy dark matter the diffusive term in~\Eq{SHE:Helmholtz} represents gravity 
		in the Newtonian approximation of the Einstein field equations, as per Sec.~\ref{sss:SNE}, 
		while the local term corresponds to the inclusion of a cosmological constant $\Lambda$ 
		in this approximation~\citep{kiessling2003jeans}.
This is necessary if one wants to account for a dark energy component to cosmology in a Newtonian approximation. It is also a means to regularise the so-called ``Jeans swindle''---the specification that \Eq{SNE:Poisson} only relates the fluctuations of density and potential around an unspecified equilibrium~\citep{BinneyTremaine1987_GalacDynBook}, 
		see Appendix~\ref{a:Jeans}.

		In the optical context \Eqs{SHE} model a system where both Kerr (local) effects and 
		thermo-optic or elastic (diffusive nonlocal) effects are important 
		(alternatively, the diffusive term in~\Eq{SHE:Helmholtz} can be used to take account of heat losses 
		at the edges of the optical sample~\citep{ghofraniha2007shocks, Faccio2016_OpticalNewtSchro}).

We therefore take the Schr{\"o}dinger-Helmholtz \Eqs{SHE}, as our model of interest as they comprise a model that is 
	physically relevant in both astrophysics and nonlinear optics, depending on the choice of dimensionality and units.
They contain as limits both the nonlinear Schr{\"o}dinger equation, about which much is known, and the Schr{\"o}dinger-Newton equations, whose relevance is starting to come to the fore. Next we discuss weak and strong turbulence in these latter models, and introduce the process 
	of dual cascade of invariants, which is a precursor to the formation of structures at the largest scale in Schr{\"o}dinger-Helmholtz systems.

\subsection{\label{ss:NLSturbulence}Turbulence in the nonlinear Schr{\"o}dinger and  Schr{\"o}dinger-Newton equations}

Turbulence in laboratory Bose-Einstein condensates~\citep{Vinen2002QTurb, barenghi2001quantized, kobayashi2007quantum, krstulovic2011dispersive, tsatsos_quantum_2016, tsubota2017numerical, muller2020abrupt} and optics~\citep{LaurieEtAl2012_1DOpticalWT, dyachenko1992optical,  picozzi2014optical} is now a well-established field, and much has been understood by using the local nonlinear Schr{\"o}dinger \Eq{NLSE}. Its dynamics is rich, with weakly nonlinear waves typically coexisting with coherent, strongly nonlinear structures. The nature of these structures depends radically on the sign of the interaction term $s$ in \Eq{NLSE}. In the defocusing (repulsive) case they include stable condensates: accumulations of particles (in the Bose-Einstein condensate case) or intensity (optics) at the largest scale, with turbulence manifesting as a collection of vortices in 2D, or a tangle of vortex lines in 3D, on which the density is zero and which carry all the circulation, propagating through the condensate~\citep{tsatsos_quantum_2016, tsubota2017numerical, nazarenko2011waveturbbook}. In the focusing (attractive) case solitons and condensates are unstable above a certain density, with localised regions of over-density collapsing and becoming singular in finite time~\citep{dyachenko1992optical, sulem1999nonlinear}.

On the other hand, turbulence in the Schr{\"o}dinger-Newton \Eqs{SNE} has only recently been investigated by direct numerical simulation in the cosmological setting~\citep{mocz2017galaxyI} and appears to contain features of both the focusing and the defocusing nonlinear Schr{\"o}dinger equation. As mentioned above, at large scales the Schr{\"o}dinger-Newton model  exhibits gravitationally-driven accretion into filaments which then become unstable and collapse into spherical haloes~\citep{mocz2019first, mocz2019galaxyII} (cf.\ collapses in the focusing nonlinear Schr{\"o}dinger model driven by the self-focusing local contact potential). However, within haloes the condensate is stable, with turbulence in an envelope surrounding the core manifesting as a dynamic tangle of reconnecting vortex lines, as in the defocusing nonlinear Schr{\"o}dinger model~\citep{mocz2017galaxyI}. This is to be expected, given that the attractive feature of the  Schr{\"o}dinger-Newton model in cosmology is that it is simultaneously  unstable to gravitational collapse and stable once those collapse event have regularised into long-lived structures, and so it should contain features of both the unstable (focusing) and stable (defocusing) versions of the nonlinear Schr{\"o}dinger model.

To understand more fully the phenomenology recently reported in the Schr{\"o}dinger-Newton \Eqs{SNE}, it is tempting to apply theoretical frameworks that have been successful in explaining various aspects of turbulence in the nonlinear Schr{\"o}dinger equation. One such theory is wave turbulence: the study of random broadband statistical ensembles of weakly interacting waves~\citep{nazarenko2011waveturbbook, zakharov1992kolmogorovbook}.
	The ``turbulent'' behaviour referred to here is the statistically steady-state condition where dynamically conserved quantities cascade through scales in the system via the interaction of waves, a process analogous to the transfer of energy in 3D classical fluid turbulence (and respectively energy and enstrophy in 2D).
	Wave turbulence theory is integral to the quantitative description of both the wave component and the evolution of the coherent components
	of the nonlinear Schr{\"o}dinger system and is relevant in three regimes: de Broglie waves propagating in the absence of a 
	condensate~\citep{dyachenko1992optical, nazarenko2011waveturbbook}, Bogoliubov acoustic waves on the background of a 
	strong condensate~\citep{dyachenko1992optical, nazarenko2011waveturbbook}, and Kelvin waves that are excited on quantized vortex lines
	in a condensate~\citep{nazarenko2011waveturbbook, lvovnazarenko2010kelvin}. If the system is focusing, then the condensate is
	modulationally unstable and vortices do no appear, so acoustic and Kelvin wave turbulence will not be realised [the gravitational-type
	nonlinearity present in \Eqs{SNE} is of focusing type and so this is the situation that is most relevant to this work].
	Nonetheless, in both focusing and defocusing systems de Broglie wave turbulence theory describes how, starting from a random ensemble of
	waves, a dual cascade simultaneously builds up the large-scale condensate while sending energy to small 
	scales~\citep{nazarenko2011waveturbbook}.
 As we will describe in Sec.~\ref{ss:dualcascade} below, this dual cascade is generic in any system of interacting waves with two quadratic dynamical invariants (particles and energy in the cases of interest here). The theory of wave turbulence thus provides a universal description of how large-scale coherent structures can arise from a random background.

	The wave turbulence of \Eqs{SHE}, the fundamental process of dual cascade, and the spectra on which such cascades can occur, 
	have already been investigated theoretically and in optics experiments 
	in the one-dimensional case~\citep{BortolozzoEtAl2009_OpticalWTCondensnLight, LaurieEtAl2012_1DOpticalWT} 
	in the large-scale and small-scale limits where the dynamical equations become Eq.~\eqref{NLSE} and Eqs.~\eqref{SNE} respectively.
	To our knowledge such a study of the wave turbulence of~\eqref{SHE} has not been made in higher dimensions. 
	We begin this study in the current work.

	Having said this, we note that Ref.~\citep{picozzi2014optical} refers to the ``optical wave turbulence'' of nonlocal systems, 
	of which the Schr{\"o}dinger-Helmholtz equations are an important example. Much of Ref.~\citep{picozzi2014optical}, and references therein,
	pertains to the dynamics of inhomogeneous systems (such as modulational instability and collapse, studied by a Vlasov equation). 
	By contrast here we are concerned with the dynamics that govern statistically homogeneous systems. 
	We comment on the difference in approaches to inhomogeneous vs.\ homogeneous systems in Appendix B.
				Furthermore, a recent paper~\citep{levkov2018gravitational} has examined the formation of large-scale structure
	 in astrophysical Bose gases obeying Eqs.~\eqref{SNE}, using a kinetic formulation which was termed``wave turbulence'' in 
	 Ref.~\citep{niemeyer2020small}. We describe the similarities and differences between Ref.~\citep{levkov2018gravitational}
	 and this work in Sec.~\ref{ss:discussion}.

\subsection{\label{ss:way}Organisation of this paper}

In this work, then, we develop the theory of wave turbulence for the Schr{\"o}dinger-Helmholtz \Eqs{SHE} in the case of fluctuations about a zero background. By taking the limits of small and large $\Lambda$ we obtain the wave turbulence of the Schr{\"o}dinger-Newton \Eqs{SNE} and also review known results of the nonlinear Schr{\"o}dinger \Eq{NLSE}. Our aim is to describe the
	fundamental dynamical processes that govern the first stages of formation of a large-scale condensate from random waves 
	in cosmology and in nonlinear optics. 
From this structure gravitational-type collapses will ensue and the phenomenology described above will develop. 

In Secs.~\ref{ss:Hamiltonian} and~\ref{ss:KE} we overview the wave turbulence theory and arrive at the wave kinetic equation that describes the evolution of the wave content of the system. Section~\ref{ss:dualcascade} describes the dual cascade of energy towards small scales and particles towards large scales in the system. In Secs.~\ref{ss:KZ} and~\ref{ss:RJ} we describe respectively the scale-free pure-flux spectra and equilibrium spectra that are formal stationary solutions of the wave kinetic equation. However in Sec.~\ref{ss:dir} we show that these stationary spectra yield the wrong directions for the fluxes of energy and particles, as compared with the directions predicted in Sec.~\ref{ss:dualcascade}. We resolve this paradox by developing a reduced model of the wave dynamics in Secs.~\ref{ss:DAM} and~\ref{ss:DAMfluxes} and using it in Sec.~\ref{s:Reconciling} to reveal the nature of the dual cascades in the nonlinear Schr{\"o}dinger and the Schr{\"o}dinger-Newton limits of the Schr{\"o}dinger-Helmholtz equations. We conclude in Sec.~\ref{s:conclusion} and suggest further directions of research incorporating wave turbulence into the study of the Schr{\"o}dinger-Helmholtz equations.

\section{\label{s:WT} Building blocks of Schr{\"o}dinger-Helmholtz wave turbulence}

In this section we overview the aspects of the wave turbulence theory that we require in our description of  turbulence in the Schr{\"o}dinger-Helmholtz model. 

\subsection{\label{ss:Hamiltonian} Hamiltonian formulation of the Schr{\"o}dinger-Helmholtz equations}

To put Schr{\"o}dinger-Helmholtz turbulence in the context of the general theory of wave turbulence we need to formulate the Schr{\"o}dinger-Helmholtz \Eqs{SHE} in Hamiltonian form. For that goal we first set the system in the periodic box $\Omega=\mathbb{T}^d_L$ and decompose variables into Fourier modes 
\begin{equation*}
\psi_\mathbf{k}(t)=\frac{1}{L^{d}}\int_\Omega \! \psi(\mathbf{x},t) e^{-i\mathbf{k} \cdot \mathbf{x} } \, d\mathbf{x}\,,
\end{equation*}
 and similarly for $V_\mathbf{k}(t)$. The dynamical equations become
\begin{subequations}
 \label{SNE_Fourier} 
   \begin{align}
		i \partial_t \psi_\mathbf{k}   -   k^2 \psi_{\mathbf{k}}   -   \sum_{1,2} V_1 \psi_2 \delta^{\mathbf{k}}_{  1   2}= 0\,,
		\\ 
		-(k_1^2 + \Lambda) V_1 = \gamma \sum_{3,4} \psi_3 \psi_4^* \delta^3_{14}\,,
	\end{align}
\end{subequations}
where $V_j=V_{\mathbf{k}_j}$, $\psi_j= \psi_{\mathbf{k}_j}$, $ \sum_{i \ldots j}= \sum_{\mathbf{k}_i , \ldots ,\mathbf{k}_j}$ and $\delta^{\mathbf{k}}_{  1   2} = \delta(\mathbf{k}-\mathbf{k}_1-\mathbf{k}_2)$
 is the Kronecker delta, equal to  unity if $\mathbf{k}=\mathbf{k}_1+\mathbf{k}_2$ and zero otherwise.\footnote{\label{foot:Jeans_outset}If 
 				we start with \Eq{SNE:Poisson}, i.e.,\ $\Lambda=0$, from the outset, then we need to set $V_{\mathbf{k}=0}=0$, 
 				which is the Jeans swindle in Fourier space. 
 				This corresponds to subtraction of the mean as in \Eq{Poisson_JeansSwindle}, i.e.,\ $\langle\rho\rangle_\Omega=0$.
 				}
 
Equations~\eqref{SNE_Fourier} can be rewritten as the canonical Hamiltonian equation 
\begin{subequations}
\label{H}
	\begin{align}
		i\partial_t \psi_\mathbf{k} = {}& \frac{\partial H}{\partial\psi_\mathbf{k}^*}\,, \qquad H=H_2 + H_4\,, 
		\\
		H_2 = {}& \sum_\mathbf{k}\omega_\mathbf{k} \psi_\mathbf{k}\psi_\mathbf{k}^*\,, 
		\\
		H_4 = {}& -\frac{1}{2} \sum_{1234} W^{12}_{\,34}\psi_1\psi_2\psi_3^*\psi_4^*\delta^{12}_{34}\ . \label{H_4}
 \end{align}
 Here the Hamiltonian $H$ is comprised of the quadratic part $H_2$, which leads to linear waves with dispersion relation $\omega_\mathbf{k}=k^2$, and the interaction Hamiltonian $H_4$ which describes four-wave coupling of the $2 \leftrightarrow 2$ type. 
The interaction coefficient $W^{12}_{\,34}$ can written in the symmetric form
	\begin{align}
		W^{12}_{\,34}={}&\frac{\gamma}{4} \left(   A_{1234} + A_{2134} + A_{1243} + A_{2143} \right)\,,\label{W1234}
		\\ 
		A_{1234} ={}& 1\Big/ \big[ ( \mathbf{k}_1 - \mathbf{k}_4 )\cdot( \mathbf{k}_3-\mathbf{k}_2 ) + \Lambda) \big]\ .\label{A1234}
	\end{align} 
If we are using the Jeans swindle from the outset (see Footnote~\ref{foot:Jeans_outset}) then the sum  in \Eq{H_4} must exclude all terms when any two wavenumbers are equal.

For completeness, we note that if we include a local cubic self-interaction term $-s\lvert\psi\rvert^2\psi$ on the right-hand side of \Eq{SNE:NLS} as well as the gravitational term then the four-wave interaction coefficient would be
\begin{equation}
	\label{W1234GPP}
	W^{12}_{\,34}=-s+\frac{\gamma}{4} \left(   A_{1234} + A_{2134} + A_{1243} + A_{2143} \right),
\end{equation}
with $A_{1234}$ as in \Eq{A1234}.
\end{subequations}
Finally, the four-wave interaction coefficient for the cubic nonlinear Schr{\"o}dinger \Eq{NLSE} is simply 
\begin{equation*}
	\label{W1234NLSE}
	W^{12}_{\,34}=-s\ . 
\end{equation*}

\subsection{\label{ss:KE}Kinetic equation and conserved quantities}

In the theory of weak wave turbulence we consider ensembles of weakly interacting waves with random phases uniformly distributed in $[0,2\pi)$, and independently distributed amplitudes~\citep{nazarenko2011waveturbbook,  choi2004probability, choi2005joint, choi2009wave}.
 We define the wave spectrum 
\begin{equation}\label{nk}
  n_\mathbf{k}=\left(\frac{L}{2\pi}\right)^d  \langle \lvert \psi_\mathbf{k} \rvert^2\rangle\,,
 \end{equation}
  where the angle brackets $\langle \ldots \rangle $ denote averaging of ``$\ldots$'' over the random phases and amplitudes.

In the limit of an infinite domain $L\to\infty$ and for weak nonlinearity $\lvert H_4/H_2 \rvert \ll1$ one can derive~\citep{dyachenko1992optical,  nazarenko2011waveturbbook, zakharov1992kolmogorovbook} a wave kinetic equation for the evolution of the spectrum. For  $2 \leftrightarrow 2$ wave processes with the interaction Hamiltonian~\eqref{H_4},  the kinetic equation is
\begin{align}\label{kin_eq}
	\begin{split}
		\partial_t n_\mathbf{k}  ={}& 4
			\pi   \!  \int  \! 
			\lvert W^{12}_{\,3\mathbf{k}} \rvert^2 \delta^{12}_{3\mathbf{k}} \delta(\omega^{12}_{3\mathbf{k}})  n_1 n_2 n_3 n_\mathbf{k}   
		\\ 
		&\times \left[\frac{1}{n_\mathbf{k}} + \frac{1}{n_3} - \frac{1}{n_1} - \frac{1}{n_2} \right]
		 	\mathrm{d}\mathbf{k}_1  \mathrm{d}\mathbf{k}_2   \mathrm{d}\mathbf{k}_3\,, 
	\end{split}
\end{align}
where $\delta^{12}_{3\mathbf{k}}$ is now a Dirac delta  function that imposes wavenumber resonance $\mathbf{k} + \mathbf{k}_3 = \mathbf{k}_1 + \mathbf{k}_2 $; likewise frequency resonance $\omega_\mathbf{k} + \omega_3 = \omega_1 + \omega_2$ is enforced by the Dirac delta $\delta(\omega^{12}_{3\mathbf{k}})$.

The kinetic equation~\eqref{kin_eq} describes the irreversible evolution of an initial wave spectrum via four-wave interaction.\footnote{
		Note that the interaction coefficient enters Eq.~\eqref{kin_eq} only through its squared modulus, 
		so that the sign of the interaction does not play a role in the weakly nonlinear limit. 
		This means that, for example, in the case of \Eq{NLSE} the buildup of a large-scale condensate via an inverse cascade 
		is the same for both the focusing and defocusing case, and the difference only enters in the strongly nonlinear evolution.
}
			 It is the central tool of wave turbulence theory at the lowest level of closure of the hierarchy of moment equations 
			 (the theory also allows the study of higher moments or even the full probability density 
			 function~\citep{nazarenko2011waveturbbook, choi2004probability, choi2005joint, choi2009wave}). 
			 Equation~\eqref{kin_eq} allows one to study the dynamical evolution of a wave spectrum from an arbitrary initial condition, 
			 provided the interaction is weak. 
			 The spectra that are of greatest interest in wave turbulence theory are the stationary solutions that we discuss in 
			 Secs.~\ref{ss:KZ} and~\ref{ss:RJ}. As well as being the first checkpoint in analysing the wave turbulence of a new system, 
			 these spectra also frequently characterise the time-dependent dynamics. 
			 We shall return to this point in Sec.~\ref{ss:discussion}.

As the spectrum evolves under the action of \Eq{kin_eq} the following two quantities are conserved by the kinetic equation
\begin{subequations} 
	\label{invariants}
	\begin{align}
		N = {}& \int \! n_\mathbf{k} \,  \mathrm{d}\mathbf{k}\,,  
		\\
		E = {}& \int \! \omega_\mathbf{k} n_\mathbf{k}  \, \mathrm{d}\mathbf{k}\ .
	\end{align}
\end{subequations}
Here $N$ is known as the (density of) waveaction, or particle number, and is conserved for all times by the original \Eqs{SHE}, and $E$ is referred to as the (density of) energy. It is the leading-order part of the total Hamiltonian, i.e.,\ $H_2$, and is only conserved by \Eqs{SHE} over timescales for which the kinetic equation~\eqref{kin_eq} is valid.

For isotropic systems such as \Eqs{SHE} we can express the conservation of invariants~\eqref{invariants} as scalar continuity equations for the waveaction
\begin{subequations} 
\label{continuity}
	\begin{equation}
		\label{N_continuity}
		\partial_t N_k^\mathrm{(1D)} + \partial_k \eta =0\,,
		\qquad 
		N_k^\mathrm{(1D)} = \mathbb{A}^{(d-1)} n_\mathbf{k} k^{d-1}\,,
	\end{equation}
and for the energy
	\begin{equation}
		\label{E_continuity}
		\partial_t E_k^\mathrm{(1D)} + \partial_k \epsilon = 0\,, 
		\qquad 
		E_k^\mathrm{(1D)} = \omega_\mathbf{k} N_k^\mathrm{(1D)}\ .
	\end{equation}
\end{subequations}
Here $\eta = \eta(k)$ and $\epsilon=\epsilon(k)$ are, respectively, the flux of waveaction and energy through the shell in Fourier space of radius $k=\lvert\mathbf{k}\rvert$. In \Eq{N_continuity} we have defined the isotropic 1-dimensional (1D) waveaction spectrum $N_k^\mathrm{(1D)}$,   where $\mathbb{A}^{(d-1)}$ is the area of a unit $(d-1)$-sphere; likewise in \Eq{E_continuity} 
$E_k^\mathrm{(1D)}$  is the isotropic 1D energy spectrum.

In the rest of this work we will consider a forced-dissipated system, with forcing in a narrow band at some scale $k_\mathrm{f}$ and dissipation at the large and small scales  $k_\mathrm{min}$ and $k_\mathrm{max}$ respectively, and assume that these scales are widely separated $k_\mathrm{min} \ll k_\mathrm{f} \ll k_\mathrm{max}$.  The interval $k_\mathrm{f} < k < k_\mathrm{max}$ is known as the direct inertial range, and  $k_\mathrm{min} < k <  k_\mathrm{f}$ is called the inverse inertial range, because of the directions that $E$ and $N$ flow through these ranges, as we describe in the next section. In this open setup the local conservation \Eqs{continuity}   will hold deep inside the inertial ranges but the global quantities $N$ and $E$ are only conserved if the rates at which they are injected match their dissipation rates. 

	We examine the open system because it allows the nonequilibrium stationary solutions of \Eq{kin_eq} to form and persist, 
	revealing the dual cascade in its purest manifestation. 
	The alternative would be to study turbulence that evolves freely from an initial condition. 
	In that case features of the stationary solutions still often characterise the evolving spectra, see Sec.~\ref{ss:discussion}.
	We leave the study of the time-evolving case to future work and here establish the forms of the stationary spectra by considering the 
	forced-dissipative setup.

\subsection{\label{ss:dualcascade} Fj{\o}rtoft argument for two conserved invariants}

The presence of two dynamical invariants $E$ and $N$ whose densities differ by a monotonic factor of $k$, here by    $\omega_\mathbf{k}=k^2$, places strong constraints on the directions in which the invariants flow through $\mathbf{k}$-space, as pointed out by Fj{\o}rtoft \citep{fjortoft1953changes}. We recapitulate his argument in its open-system form.\footnote{
	See also Chapter 4 of Ref.~\citep{nazarenko2011waveturbbook} that makes a modified argument that does not rely on the system being open.}

Consider the system in a steady state where forcing balances dissipation: at $k_\mathrm{f}$ energy and particles are injected at rate $\epsilon$ and $\eta$ respectively, and dissipated at those rates at $k_\mathrm{min}$ or $k_\mathrm{max}$.
The ratio of the density of energy to the density of particles is $k^2$, and so the energy and particle flux must be related by the same factor at all scales. At the forcing scale this means that $\epsilon \sim k_\mathrm{f}^2\eta$. 

The argument proceeds by contradiction.
Suppose that the energy is dissipated at the large scale $k_\mathrm{min}$ at the rate $\sim \epsilon$ that it is injected. Then at this scale particles would be removed at rate $\sim \epsilon/k_\mathrm{min}^2 \sim \eta k_\mathrm{f}^2/k_\mathrm{min}^2 \gg \eta$ which is impossible because then the particle dissipation rate would exceed the rate of injection. Therefore in a steady state most of the energy must be dissipated at small scales $k_\mathrm{max}$.
Likewise, if we assume the particles are removed at the small scale $k_\mathrm{max}$ at rate $\sim \eta$ then energy would be removed at the impossible rate $\sim \epsilon k_\mathrm{max}^2/k_\mathrm{f}^2\gg\epsilon$ so most of the particles must be removed at large scales $k_\mathrm{min}$ instead.

Therefore, this argument predicts that the scale containing most of the energy must move towards high $k$ while the scale containing the most particles must move towards low $k$. Particles are then removed if $k_\mathrm{min}$ represents a dissipation scale. However if there is no dissipation here then the spectrum develops a localised bump as the particles accumulate at the largest scale---this is the condensate. In this case $k_\mathrm{min}$ represents the transition scale between the condensate, which becomes strongly nonlinear as the dual cascade proceeds, and the weakly nonlinear wave component of the system which continues to obey \Eq{kin_eq}.

It is thus the Fj{\o}rtoft argument that robustly predicts that particles accumulate at the largest available scale in the system, while energy is lost by the dissipation at $k_\mathrm{max}$, a process of simultaneous nonequilibrium condensation and ``evaporative cooling''~\citep{lvov2003wave}.

The Fj{\o}rtoft argument does not specify whether the invariants move via local scale-by-scale interactions, or by a direct transfer from the intermediate to the extremal scales. In the next \Sec{ss:KZ} we consider spectra on which the two invariants move via a local cascade.

\subsection{\label{ss:KZ} Kolmogorov-Zakharov flux spectra as formal solutions of the kinetic equation}

The landmark result of the theory of weak wave turbulence is the discovery of spectra on which invariants move with constant flux through $\mathbf{k}$-space via a local scale-by-scale cascade, potentially realising the predictions of the Fj{\o}rtoft argument. [However, anticipating the results of Sec.~\ref{ss:dir}, it turns out that for the Schr{\"o}dinger-Newton \Eqs{SNE} and  nonlinear Schr{\"o}dinger \Eq{NLSE} these spectra lead in most cases to cascades with the fluxes in the wrong direction, a contradiction that we resolve in the remainder of this work.] These are the Kolmogorov-Zakharov  spectra~\citep{zakharov1992kolmogorovbook} and are analogous to Kolmogorov's famous $k^{-5/3}$ energy cascade spectrum for 3D classical strongly-nonlinear hydrodynamical turbulence~\citep{Kolmogorov1941}. 
	When they exist, they are steady nonequilibrium solutions of the kinetic equation in which the spectra are scale-invariant, i.e.,\	
\begin{equation}
\label{powerlaw_spectra}
n_\mathbf{k} \propto k^{-x}.
\end{equation}

Necessary (but not sufficient) conditions for such spectra to exist are that both the dispersion relation and interaction coefficient are themselves both scale-invariant. 
 In our case the dispersion relation is $\omega_\mathbf{k}=k^2$. 
 For the interaction coefficient we require a homogeneous function in the sense that
\begin{equation*}
	 W^{\mu\mathbf{k}_1 \, \mu\mathbf{k}_2}_{\,\mu\mathbf{k}_3 \, \mu\mathbf{k}_4} 
 		=  \mu^\beta   W^{\mathbf{k}_1\, \mathbf{k}_2}_{\,\mathbf{k}_3 \, \mathbf{k}_4}\ .
\end{equation*}
 For the Schr{\"o}dinger-Helmholtz \Eqs{SHE} we obtain a scale-invariant interaction coefficient in either the Schr{\"o}dinger-Newton limit  $\ell \ll \ell_*$ (in which case $\beta=-2$)  or in the nonlinear Schr{\"o}dinger  limit $\ell \gg \ell_*$ (where $\beta=0$). 
 
The Kolmogorov-Zakharov spectra are obtained by making a so-called Zakharov-Kraichnan transform in the kinetic equation~\eqref{kin_eq} and using the scaling behaviour of all quantities under the integral~\citep{dyachenko1992optical, nazarenko2011waveturbbook, zakharov1992kolmogorovbook}, or via dimensional analysis~\citep{nazarenko2011waveturbbook, connaughton2003dimensional}. We omit the details and quote the results here.

For systems of $2\leftrightarrow 2$ wave scattering in $d$ spatial dimensions, the spectrum that corresponds to a \emph{constant flux of particles}  and zero flux of energy has index
\begin{subequations}
	\begin{equation}
 		x_{\scriptscriptstyle \mathrm{FN}} = d + \frac{2\beta}3  - \frac23 \ . 
	\end{equation}
The spectrum of \emph{constant energy flux}  with zero particle flux is
	\begin{equation}
		x_{\scriptscriptstyle \mathrm{FE}} = d + \frac{2\beta}3\ .
	\end{equation}
\end{subequations}
In particular for the Schr{\"o}dinger-Newton \Eqs{SNE} we have $\beta=-2$, so
\begin{subequations}
\begin{align}
&x_{\scriptscriptstyle \mathrm{FN}} = 1\,,  &x_{\scriptscriptstyle \mathrm{FE}} = 5/3       &\quad 		&\mathrm{for}     &\quad    &d=3\,,  \label{KZ_indices_3d_SNE}\\
&x_{\scriptscriptstyle \mathrm{FN}} = 0 \,, &x_{\scriptscriptstyle \mathrm{FE}} = 2/3       &\quad		&\mathrm{for}     &\quad    &d=2\,, \label{KZ_indices_2d_SNE}
\end{align}
while for the nonlinear Schr{\"o}dinger \Eq{NLSE} $\beta=0$, so
\begin{align}
&x_{\scriptscriptstyle \mathrm{FN}} = 7/3\,,  &x_{\scriptscriptstyle \mathrm{FE}} = 3       &\quad 		&\mathrm{for}     &\quad    &d=3\,, \label{KZ_indices_3d_NLSE}\\
&x_{\scriptscriptstyle \mathrm{FN}} = 4/3\,, &x_{\scriptscriptstyle \mathrm{FE}} = 2       &\quad		&\mathrm{for}     &\quad    &d=2\,. \label{KZ_indices_2d_NLSE}
\end{align}
\end{subequations}
Results~\eqref{KZ_indices_3d_NLSE} and~\eqref{KZ_indices_2d_NLSE} are known~\citep{dyachenko1992optical, zakharov1992kolmogorovbook, nazarenko2011waveturbbook} but the pure-flux Kolmogorov-Zakharov spectra \Eqs{KZ_indices_3d_SNE} and~\eqref{KZ_indices_2d_SNE} for the Schr{\"o}dinger-Newton equations are new results that we report for the first time here.

\subsection{\label{ss:RJ} Equilibrium spectra}

The kinetic equation redistributes $E$ and $N$ over the degrees of freedom (wave modes) as it drives the system to thermodynamic equilibrium. Equilibrium is reached when the invariant $\sigma=(E + \mu N)/T$ is distributed evenly across all wave modes. 
This is realised by the Rayleigh-Jeans spectrum\footnote{
	Formally, achieving the Rayleigh-Jeans spectrum depends on there being a small-scale cutoff $k_\mathrm{max}$ to prevent $\sigma$ being shared over an 
	infinite number of wave modes, i.e.,\ the trivial solution $n_\mathbf{k}=0$ for every $\mathbf{k}$.
	}
\begin{equation}
\label{RJ}
n_\mathbf{k} = \frac{T}{\omega_\mathbf{k} + \mu}\,,
\end{equation}
where $T$ is the temperature and $\mu$ is the chemical potential. 

In particular the spectrum is scale invariant, satisfying \Eq{powerlaw_spectra}, when there is equipartition of 
particles only (the thermodynamic potentials $\mu,T\to\pm\infty$ such that $T/\mu=n_\mathbf{k}=\mathrm{const}$) or of 
energy only (obtained when $\mu=0$). We denote the corresponding spectral indices for thermodynamic equipartition of particles and energy, respectively, as
\begin{equation}
\label{RJ_indices}
x_{\scriptscriptstyle \mathrm{TN}}=0
\qquad \mathrm{and} \qquad
x_{\scriptscriptstyle \mathrm{TE}}=2\ .
\end{equation}

\subsection{\label{ss:dir} Directions of the energy and particle fluxes and realisability of the scale-invariant spectra}

With the various indices for the stationary Kolmogorov-Zakharov and Rayleigh-Jeans power-law spectra in hand, we now turn to the following simple argument to determine the directions of the particle and energy fluxes $\eta(x)$ and $\epsilon(x)$. 

We consider what the flux directions will be when the spectrum is a power-law as in  \Eq{powerlaw_spectra}.
	We expect the fluxes to respond to a very steep spectrum by spreading the spectrum out. Therefore for $x$ large and positive (spectrum sharply increasing towards low wavenumber) we expect both $\eta,\epsilon >0$, and for $x$ large and negative (spectrum ramping at high wavenumber) we expect $\eta,\epsilon<0$.  
	Furthermore, both fluxes will be zero for both thermal equilibrium exponents $x_{\scriptscriptstyle \mathrm{TN}}$ and $x_{\scriptscriptstyle \mathrm{TE}}$.
	Finally, the particle flux vanishes for the pure energy flux spectrum with exponent $x_{\scriptscriptstyle \mathrm{FE}}$, and the energy flux vanishes for the pure particle flux exponent $x_{\scriptscriptstyle \mathrm{FN}}$. By continuity the signs of both fluxes for all $x$ are fully determined by their signs at infinity and the locations of their zero crossings. The fluxes will schematically vary in the manner shown in 
\Fig{fig:Fluxes}(a,b) for the  Schr{\"o}dinger-Newton model  and \Fig{fig:Fluxes}(c,d) for the nonlinear Schr{\"o}dinger model. 
\begin{figure*}
     \includegraphics[width=.9\textwidth]{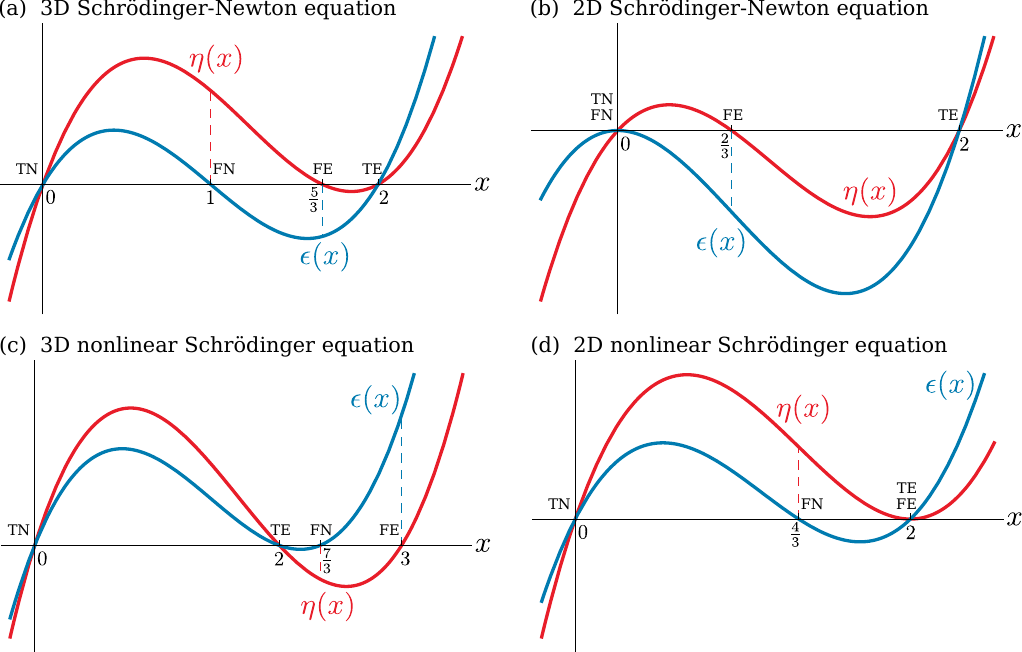}
    \caption{
    \label{fig:Fluxes}
    Particle flux $\eta(x)$ (in red) and energy flux $\epsilon(x)$ (in blue) as a function of spectral index $x$ for the  limits of the Schr{\"o}dinger-Helmholtz model. Upper panels for the Schr{\"o}dinger-Newton model in (a) $d=3$, and (b) in $d=2$. Lower panels for the nonlinear Schr{\"o}dinger model in (c)  $d=3$, and (d) $d=2$.
       Dashed lines indicate the signs of the fluxes when the spectral index takes the values $x_{\scriptscriptstyle \mathrm{FN}}$ and 			 		 	 	 	$x_{\scriptscriptstyle \mathrm{FE}}$. 
	}
\end{figure*}

First we consider the Schr{\"o}dinger-Newton equations. In both 3D and 2D at the spectral index corresponding to pure energy flux $x_{\scriptscriptstyle \mathrm{FE}}$ we find that $\epsilon$ is negative. On $x_{\scriptscriptstyle \mathrm{FN}}$, the pure particle flux spectrum, we find that $\eta$ is negative in 3D, whereas in 2D there is a degeneracy with the particle equipartition spectrum $x_{\scriptscriptstyle \mathrm{FN}}=x_{\scriptscriptstyle \mathrm{TN}}$ and correspondingly $\eta=0$ there. These findings are in contradiction to the Fj{\o}rtoft argument. 

For the nonlinear Schr{\"o}dinger equation in 3D $\epsilon$ is positive at $x_{\scriptscriptstyle \mathrm{FE}}$ and $\eta$ is negative at $x_{\scriptscriptstyle \mathrm{FN}}$. This is in agreement with the Fj{\o}rtoft argument. We therefore naively expect that in 3D the Kolmogorov-Zakharov flux cascades are possible. It turns out that the inverse particle Kolmogorov-Zakharov  spectrum is realised, with a scale-by-scale transfer of particles to small scales, however the direct energy cascade is nonlocal and the spectrum must be modified to correct a logarithmic divergence in the infrared limit, see refs.~\citep{dyachenko1992optical, nazarenko2011waveturbbook} for details. 

For the 2D nonlinear Schr{\"o}dinger equation the energy flux and equipartition spectra are degenerate $x_{\scriptscriptstyle \mathrm{FE}}=x_{\scriptscriptstyle \mathrm{TE}}$, giving $\epsilon=0$ there, and at the particle flux spectral index $x_{\scriptscriptstyle \mathrm{FN}}$ we find $\eta$ is positive.

These results for the Schr{\"o}dinger-Newton equations and 2D nonlinear Schr{\"o}dinger equation are in contradiction to the Fj{\o}rtoft argument for a forward energy cascade and inverse particle cascade. 
However the Fj{\o}rtoft argument is robust and predicts that if an initial spectrum evolves, it must push most of the energy towards small scales and particles towards large scales. We therefore conclude that the Schr{\"o}dinger-Newton \Eqs{SNE}, and the nonlinear Schr{\"o}dinger \Eq{NLSE} in 2D, do not accomplish this via the Kolmogorov-Zakharov  spectra that are determined solely by the values of the flux. To resolve this paradox we develop a simplified theory to reduce the integro-differential kinetic equation to a partial differential equation that is analytically tractable.

\subsection{\label{ss:DAM} Differential approximation model for wave turbulence}

The Kolmogorov-Zakharov solutions of the kinetic equation for the Schr{\"o}dinger-Newton equations in 3D and 2D, and for the nonlinear Schr{\"o}dinger equation in 2D, predict the wrong directions for the fluxes as compared to the Fj{\o}rtoft argument. Such solutions cannot be realised for any finite scale separation between forcing and dissipation.  From experience with other wave turbulence systems we expect that the flux-carrying spectra in these cases are instead close to the zero-flux thermal Rayleigh-Jeans solutions, but with deviations that carry the flux~\citep{nazarenko2011waveturbbook, PromentOnoratoAsinariNazarenko2012warm, connaughton2011mixed}.
These deviations are small deep inside the inertial ranges but become large at the ends, making the spectrum decay rapidly to zero near the dissipation scales.
Spectra of this sort are termed ``warm'' cascades~\citep{PromentOnoratoAsinariNazarenko2012warm, connaughton2004warm, nazarenko2006differential, boffetta2009modeling}.  A feature of these solutions is that the thermodynamic potentials $T$ and $\mu$ will be functions of the flux they have to accommodate,\footnote{
			Note that the temperature $T$ of the warm cascade refers to the energy shared between wave modes, and is not related to 	
			the temperature of particle or molecular degrees of freedom of the material at hand (Bose gas or nonlinear optical sample), which plays 	
			no role in this analysis.
} and the scales at which the spectrum decays, i.e.,
\begin{equation}
	\label{inverse_warm_relation}
	\frac{T}{\mu} = f(\eta, \omega_\mathrm{min})
\end{equation}
for the inverse cascades and 
\begin{equation}
	\label{direct_warm_relation}
	T = g(\epsilon, \omega_\mathrm{max})
\end{equation}
for  direct cascades~\citep{nazarenko2011waveturbbook, PromentOnoratoAsinariNazarenko2012warm, connaughton2011mixed}, where the functional forms of $f$ and $g$ are to be found, and we have converted from wavenumber to frequency using the dispersion relation $\omega=k^2$ (we will continue to refer to ``scales'' when discussing frequencies as the isotropy of the spectrum allows us to use the dispersion relation to convert between spatial and temporal scales).

To describe warm cascade states we develop a differential approximation model  that simplifies the kinetic equation by assuming that interactions are super-local in frequency space ($\omega_k \approx \omega_1 \approx \omega_2 \approx \omega_3$). This allows the collision integral to be reduced to a purely differential operator. Asymptotically-correct stationary solutions of this reduced model can then be found analytically, and these will be qualitatively similar to the solutions for the full kinetic equation~\citep{dyachenko1992optical, PromentOnoratoAsinariNazarenko2012warm, connaughton2011mixed, connaughton2004warm, nazarenko2006differential, boffetta2009modeling, LvovNazarenko2006differential}.

The reduction of the general four-wave kinetic equation to the differential approximation model is done explicitly  in Ref.~\citep{dyachenko1992optical}. 
Here we take a heuristic approach based on the scaling of the kinetic equation and neglect the full calculation of numerical prefactors.

We integrate over angles in $\mathbf{k}$-space and change variables to frequency.
The general form of the differential approximation model is then an ordinary differential equation  in local conservative form
\begin{subequations}\label{DAM}
\begin{equation}
	\label{DAM_time}
	\omega^{d/2-1}\frac{\partial n}{\partial t} 
	= \frac{\partial^2 R}{\partial \omega^2},
\end{equation}
where $n=n(\omega)$ is the spectrum expressed as a function of $\omega$, and the quantity
\begin{equation}
	\label{DAM_R}
	R= S \omega^\lambda n^4 \frac{\partial^2}{\partial\omega^2} \left(\frac{1}{n}\right)
\end{equation}\end{subequations}
is constructed so as to ensure that the Rayleigh-Jeans spectrum is a stationary solution  [$\partial_{\omega\omega} (1/n)$ term], the $n^4$ term derives from the fact that four-wave interactions are responsible for the spectral evolution, the total $n$ scaling matches the kinetic equation, and $S$ is a constant. 

To find $\lambda$ for the systems considered in the present work we examine how the kinetic \Eq{kin_eq} scales with $\omega$. Schematically the kinetic equation is
\begin{equation*}
	\dot{n}= \int \, W^2 n^3 \delta(\mathbf{k}) \delta(\omega) (\mathrm{d}\mathbf{k})^3 \sim n^3 k^{2\beta + 2d - 2} 
	\sim n^3 \omega^{ \beta + d - 1 }
\end{equation*} 
while the differential approximation~\eqref{DAM} scales as
\begin{equation*}
	\omega^{d/2-1}\dot{n} \sim n^3 \omega^{\lambda-4} \ .
\end{equation*}
Comparing powers of $\omega$ we find that
\begin{equation}
	\label{lambda}
	\lambda= \beta+\frac{3d}{2}+2\ .
\end{equation}

\subsection{\label{ss:DAMfluxes} Fluxes in the differential approximation}

Comparing \Eq{DAM_time} with~\eqref{N_continuity} and~\eqref{E_continuity} we see that the particle and energy fluxes  expressed as a function of $\omega$ are, up to a geometrical factor 
that can be absorbed into $S$,
\begin{equation}
	\label{DAMfluxes}
	  \eta= - \frac{\partial R}{\partial \omega}  
	  		\qquad \mathrm{and} \qquad 
	  \epsilon=- \omega  \frac{\partial R}{\partial\omega} + R\ . 
\end{equation}
respectively.

Putting  a power law spectrum $n=\omega^{-x/2}$ into \Eqs{DAM} and  \eqref{DAMfluxes} allows us to find expressions for the fluxes. 
The particle flux is 
\begin{equation*}
\eta		=  - \frac{x}{2} \left(\frac{x}{2} -1 \right)  \left( \beta + \frac{3d}{2} - \frac{3x}{2} \right)  
					S \omega^{\beta + 3d/2-3x/2-1}
\end{equation*}
and vanishes when $x=0$ or $x=2$, corresponding to the thermodynamic particle  and energy  spectral indices of \Eqs{RJ_indices}. The particle flux also vanishes when $x=d+\frac{2\beta}3$, corresponding to the energy flux spectral index $x_{\scriptstyle\mathrm{FE}}$ of \Eqs{KZ_indices_3d_SNE} to \eqref{KZ_indices_2d_NLSE}.
The energy flux is
\begin{equation}
\epsilon =  -\frac{x}{2} \left(\frac{x}{2} -1 \right)  \left(   \beta + \frac{3d}{2} - \frac{3x}{2} -1  \right) 
				  S \omega^{\beta + 3d/2-3x/2}
\end{equation}
and is again zero for the Rayleigh-Jeans spectra where $x=0$ or $x=2$, and for the constant particle flux (zero energy flux) Kolmogorov-Zakharov spectrum with $x=d + \frac{2\beta}3-\frac23$.

Thus in the differential approximation model we recover the results of {\Secs{ss:KZ} and~\ref{ss:RJ}. Furthermore, this model gives a quantitative  prediction of $\eta(x)$ and $\epsilon(x)$ for all values of $x$ (to within the limits of the super-local assumption, and the numerical determination of $S$). For example taking $S=1$ and $\omega=1$ we have the cubic functions
\begin{subequations}
	\begin{align}
		\label{eta_vs_x}
		\eta 	     ={}&  - \frac{x}{2} \left(\frac{x}{2} -1 \right)  \left( \beta + \frac{3d}{2} - \frac{3x}{2} \right)\,,
		\\
		\label{epsi_vs_x}
		\epsilon ={}& -\frac{x}{2} \left(\frac{x}{2} -1 \right)  \left(   \beta + \frac{3d}{2} - \frac{3x}{2} -1  \right)\,,
	\end{align} 
\end{subequations}
that are drawn in 
\Fig{fig:Fluxes}, with $\beta=-2$ for the Schr{\"o}dinger-Newton \Eqs{SNE} and $\beta=0$ for the nonlinear Schr{\"o}dinger \Eq{NLSE}.

\section{\label{s:Reconciling}Turbulent spectra in the  Schr{\"o}dinger-Helmholtz model}
\subsection{Reconciling with the Fj{\o}rtoft argument}   

Having established the cases in which the Kolmogorov-Zakharov spectra give either the wrong flux directions or zero fluxes for the 
Schr{\"o}dinger-Newton  and the nonlinear Schr{\"o}dinger models,
 we now seek the spectra that give the correct fluxes. To agree with the Fj{\o}rtoft argument we require a spectrum for the direct inertial range that carries the constant positive energy flux $\epsilon$ from the forcing scale $\omega_\mathrm{f}$ up to the dissipation scale $\omega_\mathrm{max}$, but carries no particles. Setting  $\eta=\partial_\omega R =0$ in eqs.~\eqref{DAMfluxes} we obtain the ordinary differential equation 
\begin{equation}
	\label{ODEeps}
	\epsilon = R = \mathrm{const} > 0
\end{equation}
in the direct inertial range.

Likewise in the inverse inertial range we require a spectrum that carries the constant negative particle flux $\eta$ from $\omega_\mathrm{f}$ to dissipate at $\omega_\mathrm{min}$, but carries zero energy. Setting $\epsilon=0$ in \Eq{DAMfluxes} we obtain $\partial_\omega R=R/\omega$ and so
\begin{equation}
	\label{ODEeta}
	\eta= - \frac{R}{\omega} = \mathrm{const} < 0
\end{equation}
 in the inverse inertial range.

We now proceed in turn through the 3D and 2D Schr{\"o}dinger-Newton equations, followed by the 2D nonlinear Schr{\"o}dinger equation,
and use \Eqs{ODEeps} and \eqref{ODEeta} to resolve the predictions from \Sec{ss:dir} that are in conflict with the Fj{\o}rtoft argument. 

(A full qualitative classification of the single-flux stationary spectra in the differential approximation model for four-wave turbulence is presented in Ref.~\citep{grebenev2020dualcascades}, based on the phase space analysis of an auxiliary dynamical system. Those general results are relevant to the systems under consideration in this paper, however here we will concentrate on the particular functional form of the flux-carrying spectra in the inertial range, and establish the relationships~\eqref{inverse_warm_relation},~\eqref{direct_warm_relation} between the thermodynamic potentials and the fluxes, in the spirit of Refs.~\citep{nazarenko2011waveturbbook, PromentOnoratoAsinariNazarenko2012warm, connaughton2011mixed}.)

\subsection{\label{ss:3D_SNE} Spectra in  the 3D Schr{\"o}dinger-Newton model}

In \Sec{ss:dir} we found that in the 3D Schr{\"o}dinger-Newton \Eqs{SNE} both the particle and the energy cascade had the wrong sign on their respective Kolmogorov-Zakharov spectra. We specialise \Eq{lambda} to this model by setting  $\beta=-2$ and $d=3$  and, following Ref.\,\citep{nazarenko2011waveturbbook}, we use the ordinary differential \Eqs{ODEeps} and~\eqref{ODEeta} to seek warm cascade solutions that carry the fluxes in the directions predicted by Fj{\o}rtoft's argument.  

\subsubsection{Warm inverse particle cascade in \\ the 3D Schr{\"o}dinger-Newton model}
\label{sss:3D_SNE_inverse}

The warm cascade is an equilibrium Rayleigh-Jeans spectrum with a small deviation. Thus we propose the spectrum
\begin{equation}
	\label{warm_spectrum}
	n=\frac{T}{\omega+\mu+\theta(\omega)}
\end{equation}
and assume that the disturbance $\theta(\omega)$ is small deep in the inverse inertial range, i.e.,\ $\omega_\mathrm{min}\ll\omega\ll\omega_\mathrm{f}$.
	We substitute this into \Eqs{DAM} and impose the constant-flux condition~\eqref{ODEeta} for the inverse cascade. Linearising with respect to the small disturbance, we obtain the equation 
\begin{equation*}
	\theta''(\omega) = - \frac{\eta}{ST^3}\frac{(\omega+\mu)^4}{\omega^{7/2}}\ .
\end{equation*}
	Integrating twice, and noting that $\lvert \eta \rvert$ is negative, yields the following expression for the deviation away from the thermal spectrum that is valid deep in the inertial range
\begin{align}
	\begin{split}
		\label{3D_SNE_inverse_theta}
		\theta(\omega) {}= &\frac{\lvert \eta \rvert}{ST^3}  
									\left(	\frac{4 \omega^{5/2} }{15}		+ 		\frac{16 \mu\omega^{3/2} }{3}	\right.	
										 	\\  
									&\left. 	-	24\mu^2 \omega^{1/2}		+		\frac{16 \mu^3 }{3 \omega^{1/2}}		+		\frac{4 \mu^4}{15 \omega^{3/2}} 
									\right) \,,
	\end{split}
\end{align}	
where we have absorbed the two integration constants by renormalising $T$ and $\mu$.	
	
	We can use~\eqref{3D_SNE_inverse_theta} to obtain a relation between the flux and thermodynamic parameters of the form~\eqref{inverse_warm_relation} via the following ``approximate matching'' argument.  We need the warm cascade spectrum to terminate at the dissipation scale $\omega_\mathrm{min}$. 
Therefore near the dissipation scale we expect  $\theta(\omega)$ to become significant, compared to the other terms in the denominator of~\eqref{warm_spectrum}, i.e.,\ we expect $\theta(\omega) \sim \omega +\mu $ near $\omega_\mathrm{min}$. 
	We put these terms into balance at $\omega_\mathrm{min}$ and assume the ordering\footnote{
				If instead we set $\omega_\mathrm{min}\sim\mu$ or $\omega_\mathrm{min}\gg\mu$ then $\theta(\omega)$ 
				would not become small for any $\omega \geq \omega_\mathrm{min}$, contradicting the assumption under which we derived 
				Eq.~\eqref{3D_SNE_inverse_theta}.
	}
 $\omega_\mathrm{min}\ll\mu$. Taking the leading term from \Eq{3D_SNE_inverse_theta}, we obtain the flux scaling
\begin{equation}
	\label{3D_SNE_inverse_warm_relation}
	 \left( \frac{T}{\mu}\right)^{3}    \sim  \frac{4}{15S}\left|\eta\right| \omega_\mathrm{min}^{3/2}\ .
\end{equation}
Of course this matching procedure is not strictly rigorous as \Eq{3D_SNE_inverse_theta} was derived for small $\theta$ and we are extending it to where $\theta$ is large. Nevertheless, we expect that the scaling relation~\eqref{3D_SNE_inverse_warm_relation} will give the correct functional relationship between the thermodynamic parameters and the flux and dissipation scale. (Results derived in a similar spirit in other systems give predictions that agree well with direct numerical simulations, see e.g., Ref.~\citep{PromentOnoratoAsinariNazarenko2012warm}.)

Now we examine the structure of the inverse cascade near the dissipation scale. Assuming that the spectrum around $(\omega-\omega_\mathrm{min})\ll\omega_\mathrm{min}$ is analytic, the condition $n(\omega_\mathrm{min})=0$ suggests that the spectrum terminates in a compact front whose leading-order behaviour is of the form
$n=A(\omega-\omega_\mathrm{min})^\sigma$. 
Again we substitute this ansatz into \Eqs{DAM}, and demand that the flux is carried all the way to the dissipation scale, i.e.,\ we impose the condition~\eqref{ODEeta}. Requiring that the flux is frequency independent fixes $A$ and $\sigma$, and we obtain the compact front solution at the dissipation scale
\begin{equation}
	\label{3D_SNE_inverse_compact}
	n=\left[  \frac{9\left|\eta\right|(\omega-\omega_\mathrm{min})^2}{10S\omega_\mathrm{min}^{7/2}}\right]^{1/3}.
\end{equation}
We shall find below that the compact front solution is nearly identical near each dissipation range in each model and dimensionality that we examine. This is because the $\sim n^3$ scaling of the spectrum in \Eqs{DAM}, and the need for the compact front to vanish at the respective dissipation scale $\omega_*$ fixes $\sigma =2/3$. The only difference will be the flux and the power of the respective $\omega_*$ in the coefficient, and the sign difference in the power law.

We note that \Eq{3D_SNE_inverse_theta}   suggests that $\theta(\omega)$ could again become large at high frequency. Arguing as above, this permits the spectrum to terminate at a compact front at frequency $\omega_+ > \omega_\mathrm{min}$. One could argue likewise for the warm direct energy cascade spectrum,  see \Eq{3D_SNE_direct_theta} below. Indeed all the warm cascade spectra discussed in this paper contain the possibility that they might be bounded by two compact fronts. We discuss this matter in Appendix~\ref{a:bimodal}.

Using the differential approximation  we have shown how the inverse cascade of particles in the 3D Schr{\"o}dinger-Newton \Eqs{SNE} is  carried by a warm cascade that closely follows a  Rayleigh-Jeans spectrum in the inertial range, with a strong deviation near the dissipation scale that gave us an approximate scaling relation between the  thermodynamic parameters and the cascade parameters. 
We also investigated the structure of the spectrum at the dissipation scale and found it to be a compact front with a $\frac23$-power law that vanishes at $\omega_\mathrm{min}$.

In the rest of this work we will use the same procedures, with the model and dimensionality under consideration giving us the appropriate $\omega$-scaling in the differential approximation, to identify similar features of the cascades. First,  we turn to the direct cascade of energy in the 3D Schr{\"o}dinger-Newton \Eqs{SNE}.

\subsubsection{\label{sss:3D_SNE_direct}Warm direct energy cascade in the 3D Schr{\"o}dinger-Newton model}

To find a direct cascade of energy for the  3D Schr{\"o}dinger-Newton equations we again use the warm cascade ansatz~\eqref{warm_spectrum} and this time impose the constant energy flux condition~\eqref{ODEeps}.
We go through the same approximate matching procedure as in \Sec{sss:3D_SNE_inverse}: we find $\theta(\omega)$ under the assumption that it is small, 
\begin{align}
	\begin{split}
		\label{3D_SNE_direct_theta}
		\theta(\omega) {}= &\frac{\epsilon}{ST^3}	
								 \left( \frac{4\omega^{3/2}}{3}		-		16\mu\omega^{1/2} \right.
									  \\  
								&\left.	+ \frac{8 \mu^2}{\omega^{1/2}}		+		\frac{16\mu^3}{15\omega^{3/2}}		+		\frac{4\mu^4}	
										{35\omega^{5/2}}
								\right)\,,
	\end{split}
\end{align}
where again we have absorbed the two integration constants into $T$ and $\mu$. 
Extending~\eqref{3D_SNE_direct_theta} towards $\omega_\mathrm{max}$ where we require it to balance the other terms in the denominator of~\eqref{warm_spectrum}, and assuming\footnote{
				For $\omega_\mathrm{max}\sim\mu$ or $\omega_\mathrm{max}\gg\mu$ there is no range of $\omega \leq \omega_\mathrm{max}$ for which $\theta(\omega)$ is small. 				
				} 
$\mu \ll \omega_\mathrm{max}$ gives a scaling relation of the type~\eqref{direct_warm_relation} 
\begin{equation}
	\label{3D_SNE_direct_warm_relation}
	T^3 \sim  \frac{4}{3S}\epsilon\omega_\mathrm{max}^{1/2}.
\end{equation}

In the immediate vicinity of $\omega_\mathrm{max}$ we again expect a compact front. Substituting $n=A(\omega_\mathrm{max}-\omega)^\sigma$ into~\eqref{ODEeps} gives the leading-order structure
\begin{equation}
	\label{3D_SNE_direct_compact}
	n=\left[  \frac{9\epsilon (\omega_\mathrm{max}-\omega)^2}{10S\omega_\mathrm{max}^{9/2}}\right]^{1/3}\ .
\end{equation}

Again we note that \Eq{3D_SNE_direct_theta} suggests that $\theta$ can be made large at some low frequency that would lead to a second compact-front cutoff at $\omega_-<\omega_\mathrm{max}$. All the warm cascade spectra we discuss here have the potential to be terminated at two compact fronts. This is discussed in Appendix~\ref{a:bimodal}.

\subsubsection{\label{sss:3D_SNE_dual} Warm dual cascade in the 3D Schr{\"o}dinger-Newton model}

In summary, the results of \Secs{sss:3D_SNE_inverse} and~\ref{sss:3D_SNE_direct} predict that for the 3D  Schr{\"o}dinger-Newton model in the forced-dissipative setup, the movement of particles to large scales and energy to small scales is realised by a dual warm cascade spectrum. This spectrum starts close to the Rayleigh-Jeans distribution~\eqref{RJ} near the forcing scale $\omega_\mathrm{f}$ and then deviates strongly away, until it vanishes at compact 2/3 power-law fronts at the dissipation scales $\omega_\mathrm{min}$ and $\omega_\mathrm{max}$. 
\begin{figure}
     \includegraphics[width=.99\columnwidth]{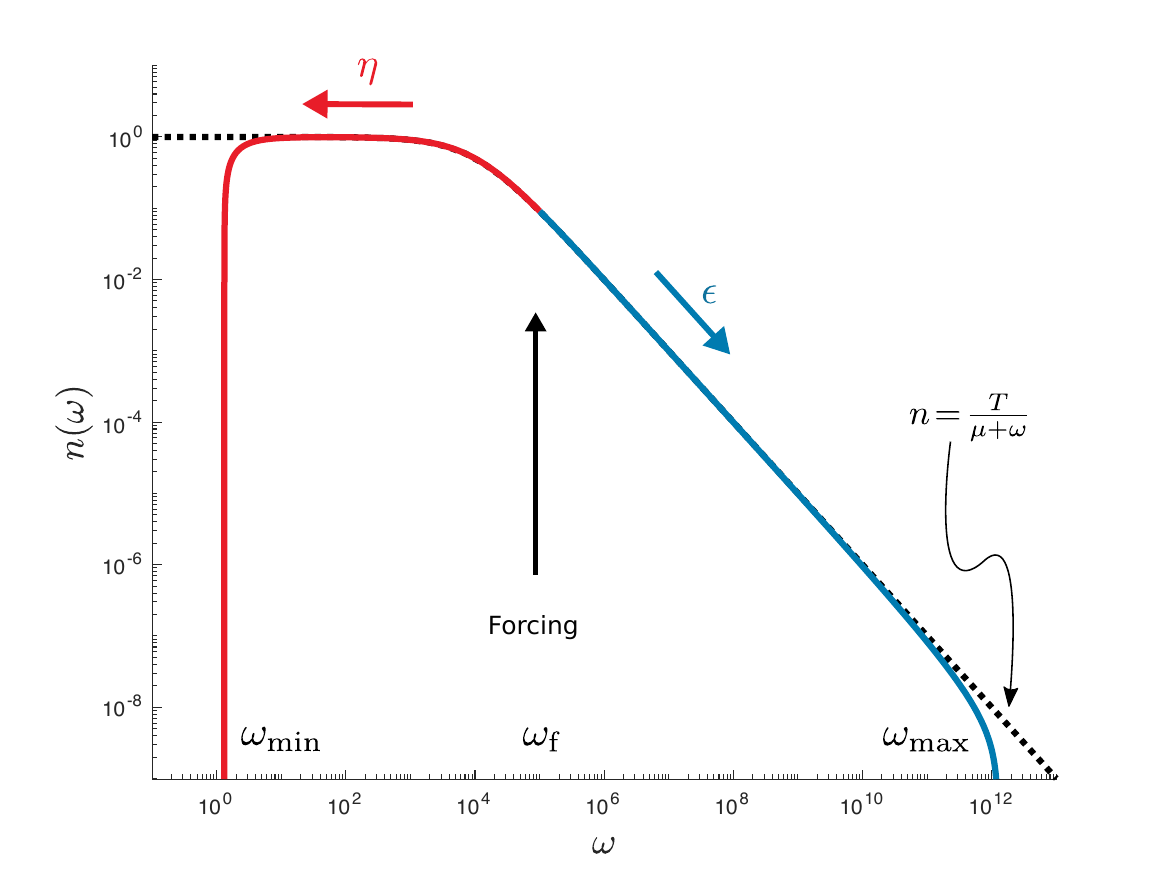}
    \caption{
    \label{fig:3D_SNE_dual}
    Dual warm cascade in the 3D Schr{\"o}dinger-Newton equations. The inverse particle cascade, with negative particle flux $\eta$, is shown in red. In blue is the direct energy cascade with positive flux $\epsilon$. The black dashed line is the thermodynamic equipartition spectrum~\eqref{RJ}. (See main text for parameters.)
	}
\end{figure}
We show the dual warm cascade in Fig.~\ref{fig:3D_SNE_dual}, which was obtained by numerically integrating Eq.~\eqref{ODEeps} forwards and Eq.~\eqref{ODEeta} backwards from the initial condition that the spectrum and its derivative matched the Rayleigh-Jeans spectrum~\eqref{RJ} at $\omega_\mathrm{f}=10^5$, with $T=\mu=10^4$. The warm cascades carry a particle flux $\eta=-3.75$ to large scales and energy flux $\epsilon=\omega_\mathrm{f}^2\lvert\eta\rvert$ to small scales, and the geometric constant $S=1$.

The dual warm cascade for the 2D Schr{\"o}dinger-Newton and 2D nonlinear Schr{\"o}dinger models can be obtained in a similar fashion. They are qualitatively similar to Fig.~\ref{fig:3D_SNE_dual} so we omit displaying them.

\subsection{\label{ss:2D_SNE} Spectra in the 2D Schr{\"o}dinger-Newton model}

Now we turn to the 2D Schr{\"o}dinger-Newton \Eqs{SNE}, setting $\beta=-2$ and $d=2$ in \Eq{lambda}.
In Sec.~\ref{ss:dir} we found that the particle equipartition and cascade spectra coincided, making the particle flux zero, and that the energy flux had the wrong sign.

\subsubsection{\label{sss:2D_SNE_inverse}Log-corrected inverse particle cascade in the~2D~Schr{\"o}dinger-Newton model}
 The degeneracy between the particle Rayleigh-Jeans and Kolmogorov-Zakharov spectra $n\sim \omega^0$  can be lifted by making a logarithmic correction to this spectrum. Substituting the trial solution 
$n=B \ln^z \!\left(\omega/\omega_\mathrm{min}\right)$
into \Eqs{DAM} and enforcing constant negative particle flux~\eqref{ODEeta} that is independent of $\omega$ gives
\begin{equation}
	\label{2D_SNE_inverse}
	n=\left[\frac{3\left|\eta\right|}S \ln   \! \left(\frac{\omega}{\omega_\mathrm{min}}\right)   \right] ^{1/3}
\end{equation}
 to leading order deep in the inverse inertial range. 

To find a relation between the thermodynamic parameters and the cascade parameters we carry out the approximate matching procedure described in Sec.~\ref{sss:3D_SNE_inverse} at low frequency $\omega\sim\omega_\mathrm{min}\ll\mu$, obtaining
\begin{equation}
	\label{2D_SNE_inverse_warm_relation}
	\left(  \frac{T}{\mu}  \right)^3\sim   \frac{\left|\eta\right|}{S}   \ln \omega_\mathrm{min}.
\end{equation}

As $\omega \to \omega_\mathrm{min}$ the spectrum in \Eq{2D_SNE_inverse} becomes zero, as we would expect given $\omega_\mathrm{min}$ is a dissipation scale. However we note that this is only a qualitative statement as subleading terms will start to dominate in this limit, meaning that~\Eq{2D_SNE_inverse} is no longer the correct stationary spectrum there.
To obtain the correct leading-order structure near $\omega_\mathrm{min}$ we look for a compact front solution and find once again a $2/3$ power law,
\begin{equation}
	\label{2D_SNE_inverse_compact}
	n=\left[  \frac{9\left|\eta\right| ( \omega-\omega_\mathrm{min}   )^2}{10S\omega_\mathrm{min}^2}   \right]^{1/3}.
\end{equation}

\subsubsection{\label{sss:2D_SNE_direct} Warm direct energy cascade in the~2D~Schr{\"o}dinger-Newton model}

To find a forward energy cascade for the 2D Schr{\"o}dinger-Newton model we again look for a warm cascade, substituting Eq.~\eqref{warm_spectrum} into~\eqref{DAM} and seeking a constant energy flux~\eqref{ODEeps}. Solving for the perturbation and matching the deviation to the other terms in the denominator in~\eqref{warm_spectrum} at $\omega \sim \omega_\mathrm{max} \gg \mu$ gives the scaling relation
\begin{equation*}
	\label{2D_SNE_direct_warm_relation}
	T^3   \sim  \frac{\epsilon\omega_\mathrm{max}^2}{6S}.
\end{equation*}
The compact front near $\omega_\mathrm{max}$ has leading-order form
\begin{equation}
\label{2D_SNE_direct_compact}
n=\left[  \frac{9\epsilon (\omega_\mathrm{max}-\omega)^2}{10 S\omega_\mathrm{max}^3}\right]^{1/3}.
\end{equation}

\subsection{\label{ss:2D_NLSE}  Spectra in the 2D nonlinear Schr{\"o}dinger  model}

In \Sec{ss:dir} we found that the Kolmogorov-Zakharov particle flux spectrum for the nonlinear Schr{\"o}dinger model was positive rather than negative, and that the Kolmogorov-Zakharov energy flux spectrum coincides with the Rayleigh-Jeans energy equipartition spectrum. 
	We specialise to the 2D nonlinear Schr{\"o}dinger \Eq{NLSE} by setting $\beta=0$ and $d=2$ in Eq.~\eqref{lambda} and take these issues in turn. (These results recapitulate and extend the discussion in Chapter 15 of Ref.~\citep{nazarenko2011waveturbbook}.)
 
\subsubsection{\label{ss:2D_NLSE_inverse} Warm inverse particle cascade in the 2D nonlinear Schr{\"o}dinger model}

The approximate matching procedure described above gives the scaling relation
\begin{equation*}
	\label{2D_NLSE_inverse_warm_relation}
	\left(  \frac{T}{\mu} \right)^3 \sim \frac{\left|\eta\right|}{6S\omega_\mathrm{min}^2}
\end{equation*}
for the inverse cascade.
The compact front solution at the dissipation scale has the structure
\begin{equation}
\label{2D_NLSE_inverse_compact}
n=\left[ \frac{9\left|\eta\right| (\omega-\omega_\mathrm{min})^2}{10S\omega_\mathrm{min}^4}\right]^{1/3}.
\end{equation}

\subsubsection{\label{ss:2D_NLSE_direct} Log-corrected direct energy cascade in the 2D nonlinear Schr{\"o}dinger model}

The degeneracy of $n\propto\omega^{-1}$ corresponding to both the Kolmogorov-Zakharov
 energy flux spectrum and the Rayleigh-Jeans energy equipartition spectrum 
 can be again lifted by making a logarithmic correction. Substituting the spectrum 
$n=(B/\omega)\ln^z \! \left(\omega_\mathrm{max}/\omega\right)$ 
 into \Eqs{DAM} and imposing \Eq{ODEeps} we obtain
\begin{equation}
	\label{2D_NLSE_direct}
	n = \frac{1}{\omega} \left[ \frac{3\epsilon\ln   ( \omega_\mathrm{max}/\omega) }{S}\right]^{1/3}\  .
\end{equation}
Comparing \Eq{2D_NLSE_direct} to the energy equipartition spectrum $n=T/\omega$ we have a relation of the kind in \Eq{direct_warm_relation}, namely
\begin{equation}
\label{2D_NLSE_direct_warm_relation}
	T^3 \sim \frac{3\epsilon}{S} \ln \omega_\mathrm{max}.
\end{equation}
We obtain the same scaling  [apart from the factor of 3 on the right-hand side of~\Eq{2D_NLSE_direct_warm_relation}]  if we assume a warm cascade and carry out the approximate matching procedure as described in Sec.~\ref{sss:3D_SNE_inverse}. This is natural as the log-corrected solution~\eqref{2D_NLSE_direct} is of a prescribed form whereas in the warm cascade argument the perturbation $\theta$ is not constrained from the outset, so the two solutions are two different perturbations from the thermal spectrum. However by continuity they should give the same scaling of thermal with cascade parameters, differing only by an $\mathcal{O}(1)$ constant. 

As in Sec.~\ref{sss:2D_SNE_inverse} the log-corrected spectrum~\eqref{2D_NLSE_direct} becomes zero at the dissipation scale. However the structure will not be correct here as sub-leading terms would start to become significant. The correct leading-order structure for the front is again the $2/3$ power-law
\begin{equation*}
	\label{2D_NLSE_direct_compact}
	n=\left[  \frac{9\epsilon(\omega_\mathrm{max}-\omega)^2 }{10S\omega_\mathrm{max}^5}\right]^{1/3}\ .
\end{equation*}

\subsection{\label{ss:3D_SHE_crossover}Crossover from warm to Kolmogorov-Zakharov cascade in the 3D Schr{\"o}dinger-Helmholtz model}

As mentioned in Sec.~\ref{ss:dir} the dual cascade in the 3D nonlinear Schr{\"o}dinger limit of~\eqref{SHE} is achieved by a scale-invariant Kolmogorov-Zakharov spectrum, rather than the warm cascade discussed in Sec.~\ref{ss:3D_SNE} for the 3D Schr{\"o}dinger-Newton limit.
Both these two regimes may be accessed if the removal of waveaction from the weakly nonlinear wave content of the system (through dissipation or absorption into the condensate) is situated at larger scales than the cosmological constant which controls the crossover between the two limits of Eqs.~\eqref{SHE}, i.e.,\ if  $\omega_\mathrm{min} \ll \Lambda$. We sketch this schematically in Fig.~\ref{fig:3D_SHE_crossover}(a) when $\omega_\mathrm{f}\gg\Lambda$, so the crossover from the Kolmogorov-Zakharov to the warm cascade happens in the inverse inertial range, and in Fig.~\ref{fig:3D_SHE_crossover}(b) when  $\omega_\mathrm{f}\ll\Lambda$ and the crossover happens in the direct inertial range.
\begin{figure*}
     \includegraphics[width=.9\textwidth]{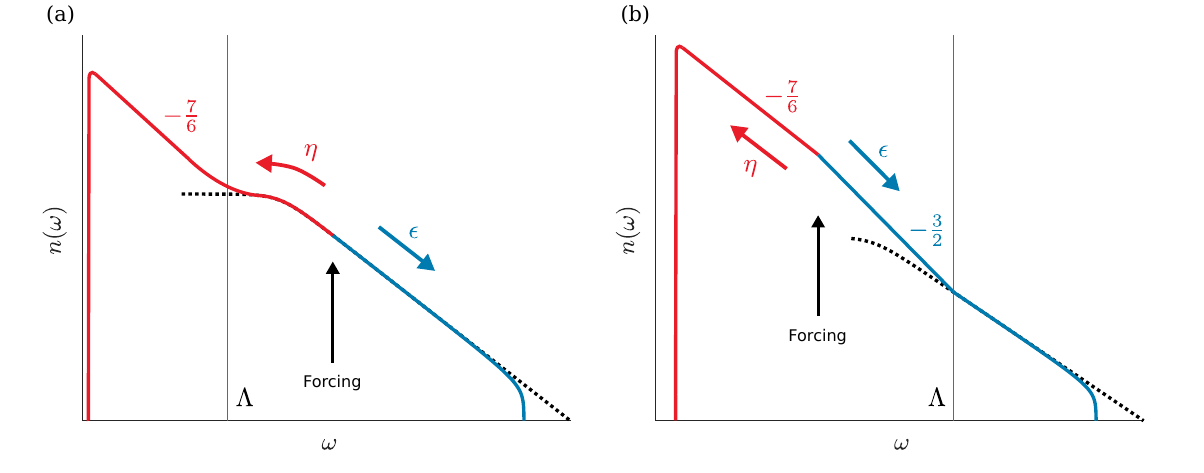}
    \caption{
    \label{fig:3D_SHE_crossover}
    Sketch of the crossover from a warm cascade [which follows closely the thermodynamic spectrum~\eqref{RJ} shown in black dashes] to a scale-free cascade, the latter with the Kolmogorov-Zakharov spectral indices shown, 
    	in the 3D Schr{\"o}dinger-Helmholtz equations.
 The crossover happens around $\omega\approx\Lambda$, with the warm cascade in $\omega\gg\Lambda$ and the Kolmogorov-Zakharov cascade in $\omega\ll\Lambda$. Depending on placement of the forcing scale the crossover happens (a) in the inverse cascade (shown in red) or (b) in the direct cascade (shown in blue).
	}
\end{figure*}

Note that Fig.~\ref{fig:3D_SHE_crossover} is a sketch and not produced directly by using the stationary differential approximation model~\eqref{DAMfluxes}. This is because in the crossover regime $\omega\approx\Lambda$ the interaction coefficient~\eqref{W1234} cannot be put into scale-invariant form. Accurate realisations of Fig.~\ref{fig:3D_SHE_crossover} must await direct numerical simulation of Eqs.~\eqref{SHE} in future work.

The crossover from a scale-invariant cascade dominated by flux  to an equipartition-like spectrum at small scales is common in turbulence, when a flux-dominated spectrum runs into a scale where the flux stagnates and thermalises. The stagnation is due to a mismatch of flux rate between the scale-invariant spectrum and the small-scale processes, whether that be (hyper-)dissipation in hydrodynamic turbulence~\citep{connaughton2004warm, frisch2008hyperviscosity}, or a different physical regime such as the crossover from hydrodynamic to Kelvin wave turbulence in superfluids~\citep{lvov2007bottlenecksuperfluid}. Our case here, the crossover from the nonlinear Schr{\"o}dinger to the Schr{\"o}dinger-Newton regime, is more like the latter but again the details await further work.

\section{\label{s:conclusion}Conclusion and outlook}
\subsection{\label{ss:discussion} Summary and discussion of results }

In this work we have developed the theory of weak wave turbulence  in the Schr{\"o}dinger-Helmholtz \Eqs{SHE}, which contain as limits both the nonlinear Schr{\"o}dinger and Schr{\"o}dinger-Newton \Eqs{NLSE} and \eqref{SNE}.
	We obtained the kinetic equation for the Schr{\"o}dinger-Helmholtz model in the case of four-wave turbulence, that is of random fluctuations of the field with no condensate present, and we used the Fj{\o}rtoft argument to predict the dual cascade of particles upscale and energy downscale in this model.

Using the differential approximation of the full kinetic equation, we have characterised the statistically steady states of its Schr{\"o}dinger-Newton  and nonlinear Schr{\"o}dinger limits in the case of a forced-dissipated system. 
	We found that the dual cascade is achieved via a warm spectrum for the Schr{\"o}dinger-Newton limit in 2D and 3D, and for the nonlinear Schr{\"o}dinger limit in 2D. For the 3D nonlinear Schr{\"o}dinger limit the Kolmogorov-Zakharov spectra are responsible for the cascades, and we have schematically illustrated the crossover between the warm and Kolmogorov-Zakharov cascades when both limits of the full Schr{\"o}dinger-Helmholtz model are accessible.
	
	Finally we found scaling relationships between the thermodynamic parameters and the fluxes and dissipation scales of the type~\eqref{inverse_warm_relation} and~\eqref{direct_warm_relation} for these cascades. We have thus characterised the processes by which particles are condensed at the largest scales, and energy sent to small scales, in both limits of the Schr{\"o}dinger-Helmholtz model. The results for the nonlinear Schr{\"o}dinger model have  appeared in the literature before, but the results for the  Schr{\"o}dinger-Newton model  are new and are relevant to the problem of cosmological structure formation in a fuzzy dark matter universe, and to optical systems where nonlocal effects are significant.
 
	For the bulk of this work we considered an open system where forcing matched dissipation. 
	This allowed us to discuss the stationary warm spectra that will realise the dual cascade in the forced-dissipated system.
	There remains the question of how the dual cascade will be realised in the time-dependent case 
	where turbulence evolves from an initial condition; such a case is far more relevant when discussing 
	the formation of galaxies, and realistic protocols in an optics experiment.
	Experience with other wave turbulence systems shows that time-dependent cascades are strongly controlled 
	by the capacity of the relevant flux spectra, defined as follows.
	We consider pushing the dissipation scales towards the extremes $k_\mathrm{min}\to 0$ and $k_\mathrm{max}\to\infty$. 
	If in this extremal case the integral defining an invariant, cf.~\eqref{invariants}, converges (or diverges) 
	at the limit towards which that invariant is cascading, then the spectrum is said to have finite (infinite) capacity, respectively. 
	It has been observed elsewhere that for finite capacity systems the cascading invariant fills out the inertial range in the wake of
	a self-similar front that reaches the dissipation scale in finite time, even in the extremal case, 
	and then reflects back towards the forcing scale, with the Kolmogorov-Zakharov flux spectrum established behind the returning front. 
	By contrast for infinite capacity systems the front takes infinite time 
	to establish in the extremal case~\citep{semikoz1995kinetics, semikoz1997condensation, galtier2000weak}.

	For the small and large-scale limits of the Schr{\"o}dinger-Helmholtz equations we have found that the flux-carrying spectra are the 
	Rayleigh-Jeans-like warm spectra, except for the nonlinear Schr{\"o}dinger limit in 3D where they are Kolmogorov-Zakharov spectra.
	It is easy to check that for all these cases, the inverse particle cascade has finite capacity and the direct cascade has infinite capacity. 
	We therefore expect that in an evolving system the inverse cascade will resemble the stationary spectra we have discussed, 
	and that these spectra will be established in finite time, but will have parameters ($\mu$ and $T$ in the warm case) that vary with time. 
	As for the direct cascade, for an unforced system there is always a $k_\mathrm{max}$ sufficiently remote that the energy in 
	any initial condition is insufficient to fill the cascade spectrum. 
	Therefore we do not expect that the direct energy cascade spectrum will be realised generically in systems evolving according to~\eqref{SHE},
	although we might expect to see the cascade when $k_\mathrm{max}$ is small enough, and the initial condition contains enough energy to
	act as a reservoir with which to fill the cascade spectrum.
	
	Our hypotheses above regarding the time-evolving case are in broad agreement with numerical results in the recent study of the 
	3D Schr{\"o}dinger-Newton equations by Levkov \emph{et al.}~\citep{levkov2018gravitational}. 
	They show by direct numerical simulation that, starting from a statistically homogeneous random field, 
	the formation of coherent structures is preceded by a kinetic evolution, 
	after which the structures become inhomogeneous due to a gravitational Jeans instability 
	(the latter collapsed structures are what they call a condensate and the condensation time they report is the time of collapse, 
	terminology we shall adhere to while comparing our study to theirs). Moreover, they argue that this kinetic evolution is governed not by 
	pure flux spectra of Kolmogorov-Zakharov type, but rather by a process of thermalisation. 
	Their conclusion entirely agrees with the scenario of large-scale structure formation via a warm cascade, 
	but the theory we have developed in this work suggests an explanation that is different from the interpretation given 
	in~\citep{levkov2018gravitational}.

	First, we can quantitatively demonstrate agreement between the wave turbulence theory of this paper and the numerics of 
	Ref.~\citep{levkov2018gravitational} 
	by estimating the characteristic timescale  $\tau_\mathrm{kin}\sim N/\lvert\mathrm{Coll}[n_\mathbf{k}]\rvert$
	over which Eq.~\eqref{kin_eq} acts, where $N$ is a typical value of the spectrum and 
	$\mathrm{Coll}[n_\mathbf{k}]$ is the right-hand side of~\eqref{kin_eq}, whose size is estimated in 	Appendix~\ref{a:inhomog}, 
	Eq.~\eqref{coll_estimate}. Taking values from the Gaussian initial spectrum of~\citep{levkov2018gravitational} 
	we obtain a characteristic kinetic timescale of $\tau_\mathrm{kin}\sim \SI{4.3e5}{}$ in dimensionless units.
	This compares favourably to the 	condensation timescale of $\SI{1.08e6}{}$ reported in~\citep{levkov2018gravitational} 
	for this initial condition: large-scale homogeneous structure forms over a timescale of  roughly $2\tau_\mathrm{kin}$
	before a gravitational instability collapses this structure into a compact object.
	This lends credence to our kinetic equation capturing the essence of the condensation processes examined by Levkov \emph{et al.}
	Furthermore they give timescale estimates for kinetic condensation in dimensional units for two models of self-gravitating bosons, 
	which links our results to astrophysically relevant processes.
	
	The points of difference between this study and Ref.~\citep{levkov2018gravitational} lie in the nature of the 
	kinetic equations that are used in each. 
	Levkov \emph{et al.} derive a Landau-type differential kinetic equation by assuming that only boson-boson interactions that are 
	strongly nonlocal 	in physical space contribute to the dynamics,	which leads to small-amplitude scattering, 
	an assumption that becomes more accurate at higher energies. They also imply that the lack of Kolmogorov-Zakharov cascades 
	is due to the nonlocality of the system.
	By contrast our kinetic equation is derived without restriction to strong nonlocality, and is valid at arbitrary energies. 
	Importantly, we attribute the lack of Kolmogorov-Zakharov cascades to the fact that 
	they give the wrong flux directions, rather than the effects of nonlocality.

	Additionally,	when we reduce our kinetic equation~\eqref{kin_eq} to the differential approximation model~\eqref{DAM},
	the latter is explicitly constructed to keep the general Rayleigh-Jeans spectrum~\eqref{RJ} as a stationary solution. 
	However the only thermodynamic spectrum that solves the differential kinetic equation in 
	Ref.~\citep{levkov2018gravitational} is the energy equipartition spectrum:~\eqref{RJ} with $\mu=0$. 
	The low-energy part of the general Rayleigh-Jeans spectrum 	is excluded from their solution, yet we argue that this is the part
	responsible for the dynamical inverse cascade of particles that builds large-scale structure.
	Despite this, at the condensation time Levkov \emph{et al.} observe excellent agreement between 
	the spectrum obtained by direct numerical simulation of~\eqref{SNE}, the spectrum obtained by evolving their kinetic equation, 
	and the low-energy part of the energy equipartition spectrum. 
	
	The agreement with the energy equipartition spectrum at the condensation time we explain by noting that $\mu=0$ is indeed 
	the criterion for condensation in 	the local Eq.~\eqref{NLSE}~\citep{connaughton2005condensation}, and the arguments are sufficiently 
	general that this criterion should apply universally. As mentioned above, we conjecture that the time-evolving spectra might be  
	Rayleigh-Jeans-like, with time-dependent thermodynamic parameters. 
	As the system evolves towards the condensation time we expect to see $\mu(t)$ shrink $0$, 
	leaving only the energy equipartition spectrum at the condensation time, as observed in~\citep{levkov2018gravitational}. 
	The deviation of the observed spectrum from the thermodynamic one at high energies might be related to the infinite capacity of the direct 
	warm cascade spectrum, meaning that the cascade may have had insufficient time to fill out at the highest frequencies, as mentioned above. 
	Indeed Levkov \emph{et al.} make reference to this part of the spectrum having a slow thermalisation timescale. 
	
	Thus, we summarise that our kinetic equation and its differential approximation is more general than that of 
	Ref.~\citep{levkov2018gravitational}, in terms of not being restricted to highly nonlocal interactions, and containing the general 
	Rayleigh-Jeans spectrum that could explain more features of the evolution in the four-wave kinetic regime.	
	Clearly further work is needed to explore and test these hypotheses.
	
\subsection{\label{ss:outlook} Outlook for wave turbulence in Schr{\"o}dinger-Helmholtz systems }
 
		We now speculate on what further perspectives wave turbulence could bring to the astrophysical and optical systems 
		to which \Eqs{SHE} apply. Focusing first on the astrophysical application, our results suggest that 
		the first process that starts to accumulate a condensate of dark matter particles at large scales in the early universe
		is an unsteady weakly nonlinear evolution that bears the hallmark of a warm dual cascade. 
Following this initial phase of condensation the subsequent evolution would follow the same broad lines as has already been documented in the literature, namely that gravitational collapse into a collection of virialised 3D spheroidal haloes will ensue~\citep{levkov2018gravitational, mocz2019first}. 

We also conjecture that wave turbulence  may have much to say regarding certain other details that have already been noted. For example, the structure of haloes has been reported as a solitonic core that is free of turbulence surrounded by a turbulent envelope~\citep{mocz2017galaxyI}. The exclusion of turbulence from the core is reminiscent of the externally trapped defocusing nonlinear Schr{\"o}dinger \Eq{NLSE}, where wave turbulence  combined with wavepacket (Wentzel-Kramers-Brillouin) analysis predicts the refraction of Bogoliubov sound waves towards the edges of the condensate, where transition from the three-wave Bogoliubov wave turbulence  to four-wave processes could occur~\citep{lvov2003wave}. On the other hand the virialisation of haloes suggests a condition of critical balance where the linear propagation and nonlinear interaction timescales of waves are equal scale by scale. In that case the weak wave turbulence described here is not applicable and new spectral relations must be found based on the critical balance hypothesis~\citep{nazarenko2011waveturbbook,  nazarenko2011critical}.

After the formation of haloes the next step of the evolution will be their mutual interaction. 
	As mentioned in Sec.~\ref{ss:NLSturbulence}, in nonlinear optics experiments and simulation of \Eqs{SNE} in one dimension
	(with six-wave interactions taken into account to break the integrability of the system), it has been observed 
 that a random field creates a condensate via the dual cascade, which then collapses into solitons. These solitons then interact via the exchange of waves and finally merge into one giant soliton that dominates the dynamics~\citep{LaurieEtAl2012_1DOpticalWT, BortolozzoEtAl2009_OpticalWTCondensnLight}. 
	It seems plausible that the same phenomenology might carry over to the Schr{\"o}dinger-Helmholtz equations, and into higher spatial dimensions.
	
Indeed, in cosmological simulations of binary and multiple halo collisions,  scattering events, inelastic collisions, and mergers are all observed~\citep{mocz2017galaxyI, bernal2006scalar, schwabe2016simulations, amin2019formation}. 
	Following such events, subsequent virialisation of the products involves ejection of some of the mass of the haloes~\citep{guzman2004evolution, guzman2006gravcooling}. 
	A detailed study of these processes should consider both the weakly nonlinear wave component and the strongly nonlinear haloes, and how the two components interact. Numerical studies could obtain effective collision kernels for those interactions in order to develop a kinetic equation for the ``gas'' of haloes that results from the collapse of a condensate. We note that work has been done in this spirit in Ref.~\citep{amin2019formation} but without detailed consideration of the wave component. In our opinion it is crucial to incorporate wave turbulence into the study of the Schr{\"o}dinger-Helmholtz model  to uncover the full richness of the behaviour that this system manifests.

	Finally, we expect that all the processes outlined above in the 3D astrophysical case---condensation via the dual cascade, fragmentation by
	modulational instability, soliton formation, and soliton interaction/merger via the exchange of weakly nonlinear waves---will be 
	qualitatively the same in 2D. This makes them all amenable to direct observation by nonlocal nonlinear optics experiments. 
	As mentioned in Sec.~\ref{sss:SNE} theoretical~\citep{navarrete2017spatial, paredes2020optics} 
	and experimental~\citep{Faccio2016_OpticalNewtSchro, Beckenstein2015_OpticalNewtSchro} comparisons have been made between 
	astrophysical phenomena and  experiments in thermo-optic media.
	To observe the wave turbulence cycle of condensation, collapse, and soliton interaction that we describe here one could also look to using 
	nematic liquid crystals and modifying the one-dimensional experiments 
	of~\citep{LaurieEtAl2012_1DOpticalWT, BortolozzoEtAl2009_OpticalWTCondensnLight} to 2D. 
	Any such experiment would need to have fine control over losses and nonlinearity strength in order to keep within the wave turbulence regime 
	while the condensate is being built up. Liquid crystals are an attractive optical medium in this respect due to several inherently tunable
	parameters~\citep{ferreira2018superfluidity} that would assist in achieving conditions relevant to wave turbulence studies.

\appendix

\section{\label{a:Jeans}Relation between the cosmological constant and the Jeans swindle}

In Sec.~\ref{sss:SHE} we motivated the inclusion of a local term in Eq.~\eqref{SHE:Helmholtz} in the dark matter application as representing a cosmological constant~\citep{kiessling2003jeans}, and asserted that this is equivalent to using the ``Jeans swindle''. In this Appendix we expand on this statement.

Eq.~\eqref{SNE:Poisson} is well-posed for spatially infinite domains in which the support of $\rho(\mathbf{x})=\lvert\psi(\mathbf{x})\rvert^2$ is compact, but if one seeks an equilibrium with spatially-constant $V$ and $\rho$ the only solution is the trivial null solution (an empty domain). 
	The Jeans swindle~\citep{BinneyTremaine1987_GalacDynBook} is the \emph{ad hoc} replacement of $V$ in~\Eq{SNE:NLS} with $\tilde{V}$ that solves 
$ \nabla^2   \tilde{V}    =  \gamma \tilde{\rho}$,
where the tildes refer to fluctuations of quantities about a nonzero equilibrium, whose existence is entirely paradoxical. 
In a periodic domain $\Omega=\mathbb{T}^d_L$ of side $L$ the equivalent problem is that \Eq{SNE:Poisson} can only be satisfied when $\Omega$ is empty, as can be seen by integrating over $\Omega$, and using the divergence theorem and the periodic boundary conditions. The Jeans swindle is then implemented by replacing~\eqref{SNE:Poisson} with
\begin{equation}
\label{Poisson_JeansSwindle}
	\nabla^2  \tilde{V}    =    \gamma \left(   \rho- \langle \rho \rangle_\Omega   \right)
\end{equation}
 where the box average of the number density $\langle\rho\rangle_\Omega  = L^{-d} \int_\Omega \rho \,\mathrm{d}\mathbf{x}$ is the equilibrium solution, and one solves only for $\tilde{V}$.

It is shown in Ref.~\citep{kiessling2003jeans} in the infinite-domain case that the Jeans swindle can formally be justified by considering the Helmholtz-like \Eq{SHE:Helmholtz} instead of \Eq{SNE:Poisson}, as the former is well-posed without the restriction of the right-hand side needing to integrate to zero, and then taking the limit $\Lambda\to0$. For the case of the periodic boundary we simply note that averaging~\eqref{SHE:Helmholtz} gives $\langle V \rangle_\Omega=-\gamma\langle \rho \rangle_\Omega/\Lambda$. Substitution back into~\eqref{SHE:Helmholtz} and writing $V = \tilde{V}+\langle V \rangle_\Omega$ recovers~\Eq{Poisson_JeansSwindle} in the limit $\Lambda\to 0$.

\section{\label{a:inhomog} Wave turbulence in inhomogeneous systems}

In the main body of this paper we have applied the theory of weak wave turbulence to the Schr{\"o}dinger-Helmholtz system, and described the initial stage of wave condensation via the dual cascade in a forced-dissipated setup. Crucial to this analysis is the assumption that the system is statistically spatially homogeneous, as only then can the dynamical variables, such as the spectrum and linear frequency, be characterised solely by time or axial distance $t$ and wavenumber $\mathbf{k}$. However for inhomogeneous systems these quantities may vary with spatial position $\mathbf{x}$. This brings into play physical effects that are not present in homogeneous systems and that are described by a different dynamical equation.
	In this Appendix we discuss the extension of wave turbulence theory to inhomogeneous systems and make simple estimates of the conditions under which the processes outlined in this paper will be the dominant dynamical processes.

To take into account inhomogeneities of the wave field we define a local spectrum that can now vary with spatial position, with characteristic spatial scale $D$, via the Wigner transform of the $\psi(\mathbf{x},t)$ field~\citep{alber1978effects}:
\begin{equation}
\label{local_spectrum}
	n_\mathbf{k}(\mathbf{x},t) = 
		\int \!   \langle \psi(\mathbf{x}-\mathbf{y}/2,t) \, \psi^*\!(\mathbf{x}+\mathbf{y}/2,t)  \rangle 
			e^{-i\mathbf{y}\cdot\mathbf{k}}   \, d\mathbf{y} \,.
\end{equation}

Let $K$ be a characteristic wavenumber associated with the spectrum. If $DK\gg1$ a Wentzel-Kramers-Brillouin analysis  gives the following Vlasov-like equation of motion for the local spectrum (see, e.g.,~\citep{alber1978effects, zakharov1985hamiltonian, dyachenko1992wave-vortex, lvov1977spatially, hall2002statistical, onorato2003landau, picozzi2007towards, levkov2018gravitational}):
\begin{equation}
\label{Vlasov-Kinetic}
	\frac{\partial n_\mathbf{k}}{\partial t}  
		+ \nabla_{\!\mathbf{k}} \tilde{\omega}_\mathbf{k} \cdot \nabla_{\!\mathbf{x}} n_\mathbf{k}
		-  \nabla_{\!\mathbf{x}} \tilde{\omega}_\mathbf{k} \cdot \nabla_{\!\mathbf{k}} n_\mathbf{k}
	= \mathrm{Coll[n_\mathbf{k}]}
	\,.
\end{equation}
The term $\mathrm{Coll}[n_\mathbf{k}]$ on the right-hand side of Eq.~\eqref{Vlasov-Kinetic} is the collision integral of the wave kinetic equation~\eqref{kin_eq} which describes spectral evolution via nonlinear wave
interactions.\footnote{
	Note
	 that in $\mathrm{Coll}[n_\mathbf{k}]$ the spectrum $n_\mathbf{k}$ is now the local spectrum defined in~\eqref{local_spectrum}. 
	 The frequency resonance condition $\delta(\omega^{12}_{3\mathbf{k}})$ can be taken between the linear frequencies of waves in the 
	 tetrad as the nonlinear frequency~\eqref{NLfreq_k} gives higher-order corrections to $\mathrm{Coll}[n_\mathbf{k}]$ that are not
	 significant during the time over which the wave kinetic equation is valid.
 }

The left-hand side of~\eqref{Vlasov-Kinetic} is the Liouville operator describing the motion of wavepackets through phase $(\mathbf{k},\mathbf{x})$ space, in which trajectories are given by Hamilton's equations. The latter are $\partial_t\mathbf{x} =\nabla_{\!\mathbf{k}} \tilde{\omega}_\mathbf{k}$ and $\partial_t\mathbf{k}=-\nabla_{\!\mathbf{x}}\tilde{\omega}_\mathbf{k}$, where the effective Hamiltonian is the renormalised dispersion relation $\tilde{\omega}_\mathbf{k}$ which, as we shall shortly discuss, is a function of the local spectrum.
	If the collision integral vanishes, wavepackets move ballistically in phase space in a manner that conserves waveaction. As they move across the inhomogeneous wave field, e.g.,\ through a turbulent patch, the amplitude of the spectrum changes and therefore $\tilde{\omega}_\mathbf{k}$ changes. In this manner wavepackets can be distorted as they move, leading to a redistribution of the spectrum and an exchange of energy between the wavepackets and the background turbulence~\citep{zakharov1985hamiltonian, dyachenko1992wave-vortex, hall2002statistical, onorato2003landau}. 

	The distortion of wavepackets brings about two effects: either wavepackets are dispersed [second term on the left-hand side of~\eqref{Vlasov-Kinetic}, noting that $\nabla_{\!\mathbf{k}} \tilde{\omega}_\mathbf{k}=\mathbf{v}_\mathrm{g}$, the group velocity], or in the case of a focusing nonlinearity such as the gravitational one considered in this paper, the wavepacket can become unstable and bunch up in physical space (third term). 
	As we justify below, these collapsing events are an incoherent version of the monochromatic modulational instability, and lead to the formation of compact strongly-nonlinear structures,  studied in deep water gravity waves in Ref.~\citep{onorato2003landau}, and in 1D local~\citep{hall2002statistical} and nonlocal optical turbulence~\citep{picozzi2011incoherent}.
		These collapses were also observed in the 3D Schr{\"o}dinger-Newton equations in their dark matter context~\citep{levkov2018gravitational}, after a period of evolution governed by four-wave kinetics, such as we describe in the main body of this paper (see Sec.~\ref{ss:discussion}). 	It is thus important to distinguish when processes associated with inhomogeneity will occur faster than processes due to four-wave interaction. Below we derive conditions to evaluate which of these two types of processes dominate the dynamics.

\subsubsection{\label{sss:renormalised_freq}Renormalised dispersion relation}

For any nonlinear equation with even-wave interactions of the type $M\leftrightarrow M$, such as the Schr{\"o}dinger-Helmholtz equations (of type $2\leftrightarrow 2$), the linear dispersion relation $\omega_\mathbf{k}$ is modified by the nonlinearity~\citep{nazarenko2011waveturbbook}. This can be seen in Eq.~\eqref{H_4}where the diagonal terms in the nonlinear Hamiltonian give a contribution whose effect is to shift the linear frequency by $\omega_{\scriptscriptstyle \mathrm{NL}}$, i.e.,\ the  dispersion relation is renormalised to
\begin{equation*}
\label{Renormfreq}
\tilde{\omega}_\mathbf{k} =\omega_\mathbf{k} + \omega_{\scriptscriptstyle \mathrm{NL}} \,.
\end{equation*}
This frequency shift is the leading effect of the nonlinearity, and does not lead to interaction between wave modes. 

For the Schr{\"o}dinger-Helmholtz equations the nonlinear frequency correction is
\begin{equation}
\label{NLfreq_k}
\omega_{\scriptscriptstyle \mathrm{NL}}(\mathbf{k})
	=
	-\gamma \sum_{\mathbf{k}_1}
		\left(			\frac{1}{\lvert \mathbf{k}-\mathbf{k}_1 \rvert^2 + \Lambda} + \frac{1}{\Lambda}		\right)
		\lvert \psi_1 \rvert^2
\end{equation}
and depends on both $\mathbf{k}$ and the spectrum. When the spectrum is spatially dependent, as in~\eqref{local_spectrum}, then the renormalised frequency also varies in space, leading to the distortion of wavepackets described above. 

We can conveniently estimate the size of $\omega_{\scriptscriptstyle \mathrm{NL}}$ in the case of weak inhomogeneity $DK\gg1$. Then in physical space Eq.~\eqref{NLfreq_k} is replaced by	
\begin{equation}
\label{NLfreq_nonlocal}
 	\omega_{\scriptscriptstyle \mathrm{NL}}(\mathbf{x},t) =  
 		-\gamma\int \!  G_{\ell_*}\!(\mathbf{x}-\mathbf{y}) \mathcal{N}(\mathbf{y},t)\, d\mathbf{y} 
\end{equation}
\citep{picozzi2007towards},
where $\mathcal{N}(\mathbf{x},t)=(2\pi)^{-d}\int\! n_\mathbf{k}(\mathbf{x},t)\,d\mathbf{k}=\langle\lvert\psi(\mathbf{x},t)\rvert^2\rangle$ is the average local level of fluctuations, whose typical amplitude we denote $N$. Here $G_{\ell_*}\!(\mathbf{x}-\mathbf{y})$ is  the Green's function for Eq.~\eqref{SHE:Helmholtz}.
	It is useful to extract the explicit dependence on $\ell_*=1/\sqrt{\Lambda}$ by the scaling space as $\mathbf{x}=\ell_* \bm{\xi}$ and defining $G(\bm{\xi})$ as the normalised Green's function satisfying
$(\nabla^2_{\bm{\xi}} -1)G(\bm{\xi}) = \delta^{(d)}(\bm{\xi})$, and that integrates to unity. 
	Doing so we find  $G(\bm{\xi})=\ell_*^{d-2}G_{\ell_*}(\mathbf{x})$.
	(For self-consistency, passing to the local limit requires that $\tilde{\gamma}=\gamma\ell_*^2 \to \mathrm{const}$.)
	We approximate the convolution in~\eqref{NLfreq_nonlocal} by multiplying the average fluctuation level $N$ with the volume of the $d$-ball of size $\ell_*$. Neglecting geometrical factors and the sign we obtain
\begin{equation}
\label{NLfreq_measure}
	\omega_{\scriptscriptstyle \mathrm{NL}} \sim 2\gamma \ell_*^2 N.
\end{equation}

For $l_*K\ll1$ Eq.~\eqref{NLfreq_k} reduces to the well-known value for the nonlinear Schr{\"o}dinger equation $-2\tilde{\gamma}\sum_\mathbf{k}\langle\lvert \psi_\mathbf{k} \rvert^2\rangle$, so the estimate in~\eqref{NLfreq_measure} becomes exact.

	We now provide estimates on the various terms in Eq.~\eqref{Vlasov-Kinetic} in order to determine when the wavepacket collapse due to inhomogeneity will dominate over either dispersion, or four-wave nonlinear interactions.

\subsection{\label{ss:MI_condition}Incoherent modulational instability}

	 We assume an isotropic spectrum that has spectral width $\Delta K$ about the representative wavenumber $K$. Thus, in terms of the measure of the average fluctuations $N$ and neglecting geometric factors, the spectrum can be estimated as 
\begin{equation}
\label{spectrum_measure}
	n_\mathbf{k} \sim \frac{N}{K^{d-1}\Delta K}
	\,.
\end{equation}

With estimates~\eqref{NLfreq_measure} and~\eqref{spectrum_measure} we can estimate the sizes of the second and third terms on the left-hand side of~\eqref{Vlasov-Kinetic}. The second term describes the dispersion of wavepackets, which is a stabilising process. Noting that the linear frequency $\omega_\mathbf{k}=k^2$, that $\omega_{\scriptscriptstyle \mathrm{NL}}$ is $\mathbf{k}$-independent, and that the inhomogeneity of the spectrum has characteristic size $D$, we estimate
\begin{equation}
\label{dispersion_measure}
	\nabla_{\!\mathbf{k}} \tilde{\omega}_\mathbf{k} \cdot \nabla_{\!\mathbf{x}}n_\mathbf{k} \sim \frac{N}{DK^{d-2}\Delta K}
	\,.
\end{equation}

Turning to the focusing term, we note that $\nabla_{\!\mathbf{x}} \tilde{\omega}_\mathbf{k}=\nabla_{\!\mathbf{x}} \omega_{\scriptscriptstyle\mathrm{NL}}$ as the linear frequency is $\mathbf{x}$-independent. The spectrum varies over length $D$, however the convolution in~\eqref{NLfreq_nonlocal} ``smears out'' the variations of the spectrum over the length $\ell_*$, meaning that $\omega_{\scriptscriptstyle\mathrm{NL}}$ will vary over a length $\max(D,\ell_*)$. Additionally the spectrum has a $\mathbf{k}$-space width of $\Delta K$ by assumption, so we can approximate the gradient in  $\nabla_{\!\mathbf{k}}n_\mathbf{k}$ by $1/\Delta K$. Bringing these considerations together, we estimate the focusing term as
\begin{equation}
\label{focusing_measure}
	\nabla_{\!\mathbf{x}} \tilde{\omega}_\mathbf{k} \cdot \nabla_{\!\mathbf{k}}n_\mathbf{k} 
	\sim 
		\frac{\gamma\ell_*^2 N^2}{\max(D,\ell_*) K^{d-1}(\Delta K)^2}
		\,.
\end{equation}

Comparing~\eqref{dispersion_measure} and~\eqref{focusing_measure} we find that wavepacket collapse into incoherent solitons is favoured over wave dispersion when
\begin{equation}
\label{collapse_condition}
	\frac{D}{\max(D,\ell_*)}\frac{\tilde{\gamma} N}{K\Delta K}>1
	\,.
\end{equation}

To justify the assertion that these collapse events are the result of an incoherent modulational, (or Benjamin-Feir) instability, we note that the latter has been extensively studied in the 1D local nonlinear Schr{\"o}dinger equation, for example in the context of extreme ocean waves~\citep{onorato2013rogue}. In the oceanic literature an important dimensionless number has been identified that controls the tendency for polychromatic wavepackets to destabilise and form strongly nonlinear structures such as rogue waves: the Benjamin-Feir Index (BFI)~\citep{onorato2003landau, janssen2003nonlinear}.
	 In the notation we have established in this Appendix, this is
\begin{equation*}
BFI = \sqrt{\frac{\tilde{\gamma}N}{K^2}} \,,
\end{equation*}
with the modulational instability triggering the formation of nonlinear structures when $BFI>1$~\citep{janssen2003nonlinear}.
 However our condition~\eqref{collapse_condition} for inhomogenity on scales greater than the nonlocality length ($D > \ell_*$) and with a spectrum whose width is of the same order as the characteristic wavenumber ($\Delta K \sim K$) is just $BFI^2 > 1$ for wavepacket collapse. Thus we conclude that the ratio on the left-hand side of~\eqref{collapse_condition} contains the same physics as the BFI, so we identify the third term in Eq.~\eqref{Vlasov-Kinetic} with modulational instability and wavepacket collapse. Note that~\eqref{collapse_condition} is valid for both local and nonlocal nonlinearities, indicating that our condition is a generalisation of the BFI to the nonlocal case, and for the spectra of arbitrary width $\Delta K$.

\subsection{\label{ss:kinetic_condition}Kinetic regime}

The derivation of the wave kinetic equation requires that the nonlinearity is small in the original equation of motion. Quantitatively this means that the linear wave period is much smaller than the characteristic timescale for  nonlinear evolution~\citep{nazarenko2011waveturbbook}, or in other words $\lvert \omega_\mathbf{k}n_\mathbf{k} \rvert \gg \lvert \mathrm{Coll}[n_\mathbf{k}] \rvert $. We estimate the collision integral as
\begin{equation}
\label{coll_estimate}
	\mathrm{Coll}[n_\mathbf{k}] \sim \frac{\gamma^2 N^3}{(K^2+1/\ell_*^2)^2 K^{d+2}} \,,
\end{equation}
giving the first condition for wave turbulence
\begin{equation}
\label{WT_condition}
	(\gamma N)^2 \ll \frac{(K^2+1/\ell_*^2)^2 K^5}{\Delta K} \,.
\end{equation}

In inhomogeneous domains, wave turbulence processes will only dominate if in Eq.~\eqref{Vlasov-Kinetic} the collision integral is larger than the focusing term that leads to wavepacket collapse, i.e.,\ $\lvert \mathrm{Coll}[n_\mathbf{k}] \rvert > \lvert \nabla_{\!\mathbf{x}} \tilde{\omega}_\mathbf{k} \cdot \nabla_{\!\mathbf{k}}n_\mathbf{k} \rvert$. This gives a second condition for wave turbulence when the kinetic regime dominates over wavepacket collapse:
\begin{equation}
\label{kineticggMI_condition}
	\frac{1}{\max(D,\ell_*)}\frac{\ell_*^2 (K^2+1/\ell_*^2)^2 K^3}{ \gamma N (\Delta K)^2} < 1 \,.
\end{equation}

We now examine what our conditions~\eqref{collapse_condition},~\eqref{WT_condition} and~\eqref{kineticggMI_condition} imply about the applicability of the kinetic regime.

\subsubsection{\label{sss:local_limit}Kinetic regime in the local limit}

For Eq.~\eqref{Vlasov-Kinetic} to be valid, and for the nonlinearity to be local, we have $D\gg 1/K \gg \ell_*$. In this limit, condition~\eqref{WT_condition} becomes $(\tilde{\gamma}N)^2 \ll K^5/\Delta K$, and the condition for the modulational instability to be stable [i.e.,~\eqref{collapse_condition} with the ordering reversed] is $\tilde{\gamma}N < K(\Delta K)$. If the spectrum is broad ($\Delta K \sim K$) then the conditions for weak nonlinearity and modulational stability become identical, $(\tilde{\gamma}N)^2 \ll K^4$, which agrees with the physical intuition that for weak wave turbulence processes one must not have the spectrum collapsing into strongly nonlinear objects.

In the local limit condition~\eqref{kineticggMI_condition} becomes $K^3/D(\Delta K)^2 < \tilde{\gamma}N$. Thus for wave turbulence to be the dominant process the nonlinearity must satisfy the double inequality
\begin{equation}
\label{local_double_ineq}
\frac{K^3}{D (\Delta K)^2} < \tilde{\gamma}N < \min \! \left( K(\Delta K),   \frac{K^{5/2}}{(\Delta K)^{1/2}}\right) \,.
\end{equation}
These inequalities are violated when either the spectrum is too narrow, or the inhomogeneity length is too short. If $\Delta K$ is small then no matter how large $D$ is, the system is still vulnerable to modulational instability via the first term on the right-hand side of \eqref{local_double_ineq}, and if the spectrum is broad but $D\to 1/K$ the system again becomes modulationally unstable from the left-hand side of \eqref{local_double_ineq}, as both sides of the inequality approach the same value. 

We thus conclude that wave turbulence in the Schr{\"o}dinger-Helmholtz equations in their local limit requires that the spectrum is sufficiently broad everywhere, \emph{and} that the inhomogeneity length of the spectrum is sufficiently long.

\subsubsection{\label{sss:nonlocal_limit}Kinetic regime in the nonlocal limit}

Validity of~\eqref{Vlasov-Kinetic} and a nonlocal nonlinearity both require $D, \ell_* \gg 1/K$. We now consider the ordering $\ell_*\sim D$ or $\ell_*\gg D$. These give $K^6/(\Delta K)^2 \ll \gamma N$ for condition~\eqref{kineticggMI_condition}, whereas condition~\eqref{WT_condition} becomes $(\gamma N)^2\ll K^9/\Delta K$. For a broad spectrum this means that $\gamma N$ is both much smaller and much greater than $K^8$, which is impossible (the violation of the inequalities is worse for a narrow spectrum).

We conclude that nonlocal wave turbulence is not possible when $\ell_*\sim D$ or $\ell_*\gg D$. Instead, The ordering $D\gg\ell_*\gg1/K$ allows for wave turbulence for sufficiently large $D$.

\section{\label{a:bimodal}Bimodal cascade spectra in the differential approximation model}

In \Sec{ss:3D_SNE} we noted that \Eqs{3D_SNE_inverse_theta} and~\eqref{3D_SNE_direct_theta} permitted the deviation  away from the Rayleigh-Jeans spectrum $\theta(\omega)$ to become large at both low and high frequencies  for both the inverse and direct cascades. This led to the intriguing possibility that we could have a warm inverse cascade spectrum, which carries only particles, arising from the cutoff at $\omega_+$, becoming large at intermediate $\omega$ and terminating at the cutoff at $\omega_\mathrm{min}$. Similarly one could imagine that the warm direct cascade of energy might exist between $\omega_-$ and $\omega_\mathrm{max}$, terminating at compact fronts at those  frequencies. For both of these to be realised the combined spectrum would have two maxima. The frequency where the two cascades met would then be the forcing scale, i.e.,\ $\omega_\mathrm{f} = \omega_+=\omega_-$, and the forcing would be such that all the particles were swept upscale and the energy downscale with the spectrum at $\omega_\mathrm{f}$ vanishing.
 This scenario is illustrated in Fig.~\ref{fig:3D_SNE_double_peak}. 
\begin{figure}
     \includegraphics[width=.99\columnwidth]{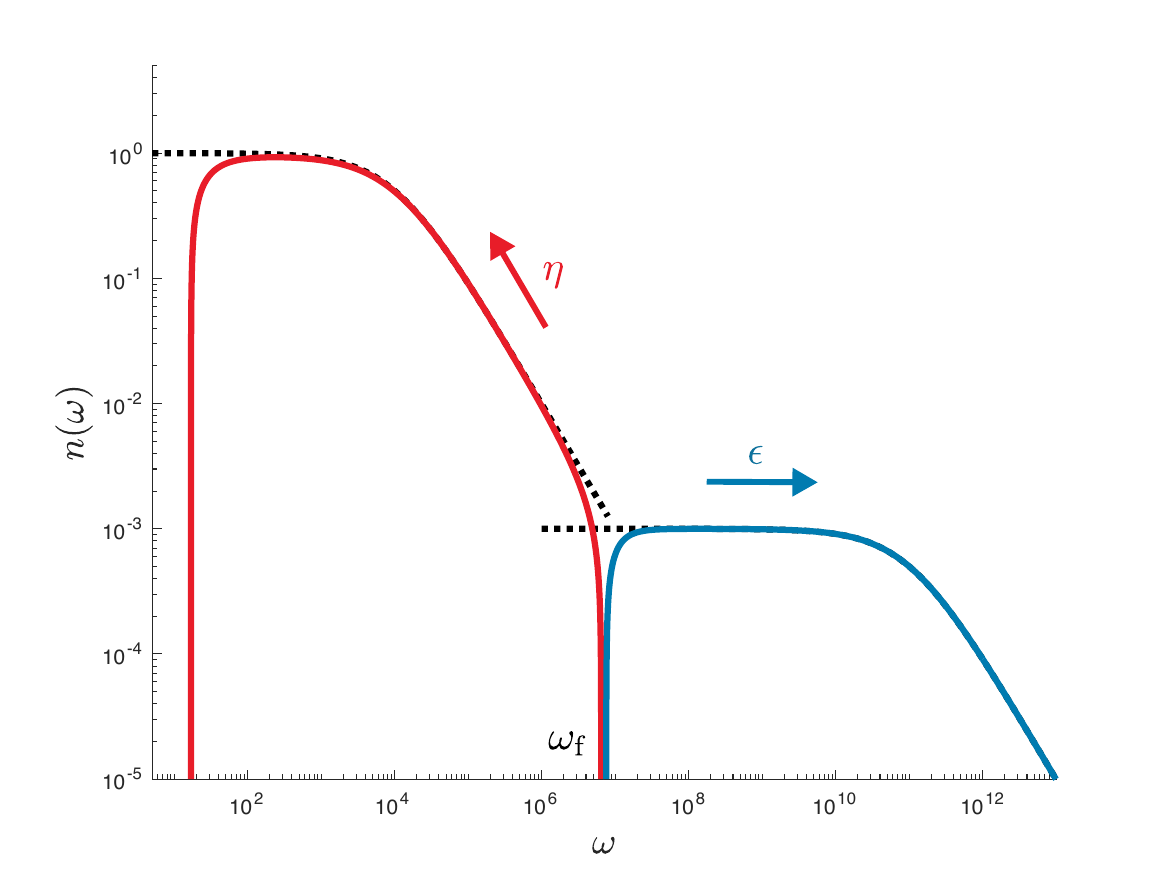}
    \caption{
    \label{fig:3D_SNE_double_peak}
        Double-peaked spectra representing a solution of the differential approximation model where particles are swept upscale (red curve) and energy downscale (blue curve) from forcing at $\omega_\mathrm{f}$ at a zero value for the spectrum.  Black dashed lines represent the two different thermodynamic spectra that match the middle of the two peaks of the spectrum. For parameters see main text.
    }
\end{figure}

	[To obtain the direct cascade, shown in blue in Fig.~\ref{fig:3D_SNE_double_peak}, we have integrated~\eqref{ODEeps} forwards to $\omega_\mathrm{max}$ and backwards to $\omega_-$ from a spectrum and its derivative matching Eq.~\eqref{RJ} at $\omega=10^{12}$ with $T=10^8$ and $\mu=10^{11}$. Likewise to obtain the inverse cascade shown in red, we integrated~\eqref{ODEeta} backwards to $\omega_\mathrm{min}$ and forwards to $\omega_+$ from a spectrum and its derivative matching Eq.~\eqref{RJ} at $\omega=10^5$ with $T=\mu=10^4$. The fluxes were $\eta=-150$ and $\epsilon=\omega_\mathrm{f}^2\lvert\eta\rvert$ with $\omega_\mathrm{f}\approx\omega_-\approx\omega_+=6.44\times10^6$. In Fig.~\ref{fig:3D_SNE_double_peak} we have chosen parameters to slightly separate $\omega_-$ and $\omega_+$ for clarity.]

We argue here that this scenario, although technically possible within the differential approximation, is implausible for more realistic models like the wave kinetic equation~\eqref{kin_eq} or the original dynamical equation itself [the Schr{\"o}dinger-Helmholtz system~\eqref{SHE} or its limits]. Note that this possibility is common to all the warm cascade spectra we discuss here. Following on from \Sec{ss:3D_SNE} we take the concrete example of the Schr{\"o}dinger-Newton model in 3D, but similar arguments can be made for either the Schr{\"o}dinger-Newton model or nonlinear Schr{\"o}dinger  model in 2D. The argument proceeds by seeking compatibility with wide inertial ranges, $\omega_\mathrm{min}\ll\omega_+$ for the inverse cascade and $\omega_- \ll \omega_\mathrm{max}$ for the direct cascade.

First considering the inverse particle cascade,  \Eq{3D_SNE_inverse_theta} and the requirement that there exists a range of $\omega>\omega_\mathrm{min}$ for which $\theta(\omega)$ is small, gives the ordering $\omega_\mathrm{min}\ll\mu$. This ordering gave the relation~\eqref{3D_SNE_inverse_warm_relation} between flux and thermodynamic parameters. Now, a cutoff at $\omega_+$ implies that in that vicinity the deviation must become comparable to the other terms in the denominator of the warm spectrum~\eqref{warm_spectrum}. We set $\theta(\omega_+)\sim \mu +\omega_+$ here. If we then let either $\omega_+ \sim \mu$ or $\omega_+\ll \mu$ and substitute~\eqref{3D_SNE_inverse_warm_relation} then we obtain $\omega_+\sim\omega_\mathrm{min}$, which is not compatible with a wide inertial range. A scale separation between forcing and dissipation is only possible if the we have the ordering $\omega_\mathrm{min}\ll\mu\ll\omega_+$ for the inverse cascade.

Next we consider the direct energy cascade. \Eq{3D_SNE_direct_theta} for the deviation, and the requirement that it must be small for some $\omega<\omega_\mathrm{max}$ gives the ordering $\mu\ll\omega_\mathrm{max}$. From this we obtained the relation~\eqref{3D_SNE_direct_warm_relation}. If we have a low-frequency cutoff at $\omega_-$ then near there it must match the other terms in the denominator of~\eqref{warm_spectrum}. We set $\theta(\omega_-)\sim \mu +\omega_-$ and consider $\omega_- \sim\mu$ and $\omega_-\gg\mu$. Substituting~\eqref{3D_SNE_direct_warm_relation} gives $\omega_-\sim\omega_\mathrm{max}$ for these two cases, which is not compatible with a wide inertial range. Therefore for the direct cascade we must have $\omega_- \ll\mu\ll\omega_\mathrm{max}$. 

Thus if we seek a double-peaked ``flux-sweeping'' spectrum with the inverse and direct warm cascades joining at $\omega_\mathrm{f}$ and the spectrum being zero there, then the cascades could not share the same thermodynamic parameters, as $\mu$ must lie deep within the inertial ranges of both cascades. Indeed, to realise such a spectrum in Fig.~\ref{fig:3D_SNE_double_peak} we have had to choose very different sets of thermodynamic parameters for each inertial range. This is technically possible within the differential approximation, as each steady cascade is described by a second order ordinary differential \Eq{ODEeps} or \Eq{ODEeta}, which only requires for its solution the value of the spectrum and its derivative at the forcing scale. 

However when considering a fuller model one must consider a more realistic forcing protocol, for example in simulations setting the spectrum to be drawn from a particular distribution at a certain level in a narrow range around  $\omega_\mathrm{f}$ at each timestep. This sets the amplitude and derivative of the spectrum at the forcing scale at the same prescribed value for both cascades, corresponding to prescription of the thermodynamic parameters $T$ and $\mu$ that both cascades share. It is therefore hard to imagine a scenario of forcing which could realise the double-peaked spectrum in a more realistic model like \Eqs{SHE} or \Eq{kin_eq}. For example the four-wave collision integral in \Eq{kin_eq} has the effect of smoothing out irregularities in the spectrum, and so we expect that any stationary solution will be at least continuous and differentiable. In this respect, this discussion stands as a cautionary example that the differential approximation includes exotic solutions like the double-peaked spectrum of Fig.~\ref{fig:3D_SNE_double_peak}, that a more physically relevant model would not permit.

\begin{acknowledgments}
SN and JS thank Marc Brachet for useful discussions regarding the Jeans swindle and its resolution. 
SN is supported by the Chaire D’Excellence IDEX (Initiative of Excellence) awarded by Universit{\'e} de la C{\^o}te d'Azur, France, the EU Horizon 2020 research and innovation programme under the grant agreements No 823937 in the framework of Marie Skodowska-Curie HALT project and No 820392 in the FET Flagships PhoQuS project and the Simons Foundation Collaboration grant `Wave Turbulence' (Award ID 651471).
JS is supported by the UK EPSRC through Grant No EP/I01358X/1.
\end{acknowledgments}

\bibliographystyle{unsrt}
\bibliography{SHE_WTbib}

\end{document}